\documentclass[11pt]{article}

\usepackage{authblk}
\usepackage{fullpage}
\usepackage{mathptmx}
\usepackage{amssymb,amsfonts,amsmath}
\usepackage[pdftex]{graphicx}
\usepackage{natbib}
\usepackage[lined,ruled,linesnumbered]{algorithm2e}
\usepackage[export]{adjustbox}
\usepackage{float}
\usepackage{booktabs}
\usepackage{framed}
\usepackage{color}
\usepackage{url}
\usepackage[implicit=false,
			colorlinks = true,
            linkcolor = blue,
            urlcolor  = blue,
            citecolor = blue,
            anchorcolor = blue]{hyperref}
\usepackage{pdfpages}

\date{}

\begin{document}

\thispagestyle{empty}
\includepdf{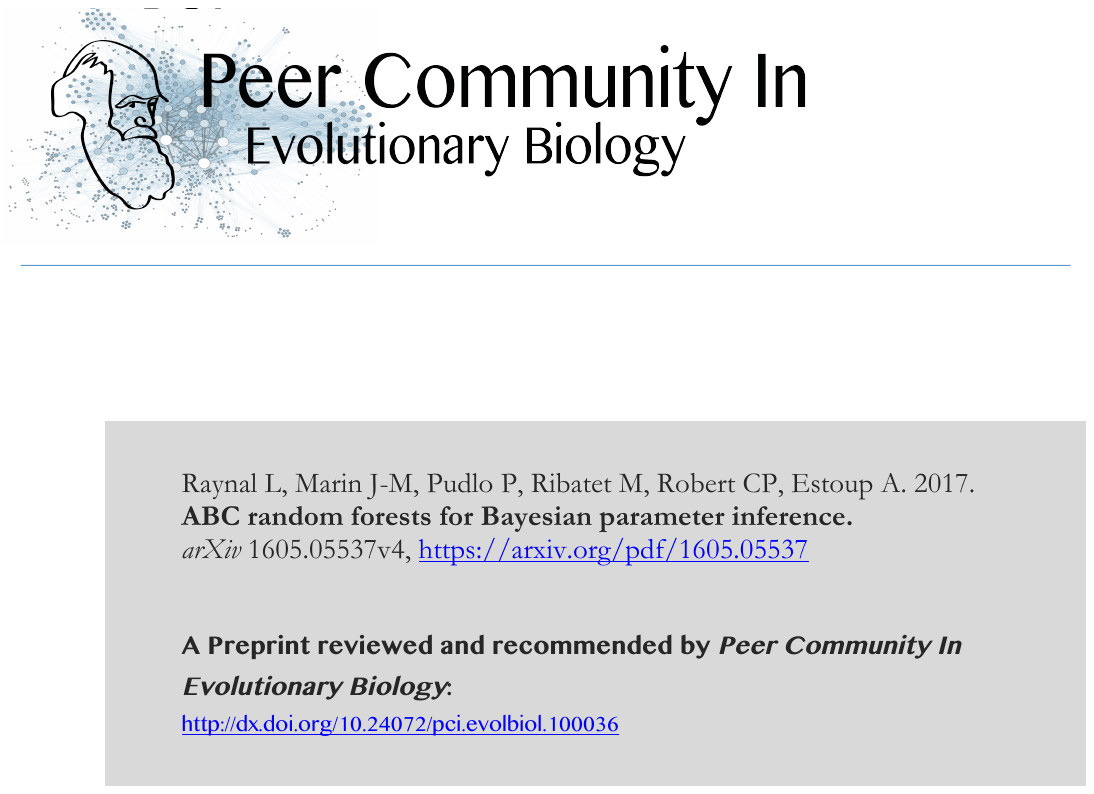}

\setcounter{page}{1}

\title{ABC random forests for Bayesian parameter inference}

\author[1]{Louis Raynal}

\author[1,6]{Jean-Michel Marin\thanks{jean-michel.marin@umontpellier.fr}}

\author[2,6]{Pierre Pudlo}

\author[1]{Mathieu Ribatet}

\author[3,4]{Christian P. Robert}

\author[5,6]{Arnaud Estoup}

\affil[1]{IMAG, Univ. Montpellier, CNRS, Montpellier, France}

\affil[2]{Institut de Math\'ematiques de Marseille, Aix-Marseille Universit\'e, France}

\affil[3]{Universit\'e Paris Dauphine, PSL Research University, Paris, France}

\affil[4]{Departement of Statistics, University of Warwick, Coventry, UK} 

\affil[5]{CBGP, INRA, CIRAD, IRD, Montpellier SupAgro, Univ. Montpellier, Montpellier, France}

\affil[6]{IBC, Univ. Montpellier, CNRS, Montpellier, France}

\maketitle

\begin{abstract}
\urlstyle{rm}
\emph{This preprint has been reviewed and recommended by Peer Community In Evolutionary Biology (\url{http://dx.doi.org/10.24072/pci.evolbiol.100036}).}
Approximate Bayesian computation (ABC) has grown
into a standard methodology that manages Bayesian inference for models associated
with intractable likelihood functions. Most ABC implementations require the
preliminary selection of a vector of informative statistics summarizing raw data.
Furthermore, in almost all existing implementations, the tolerance level that
separates acceptance from rejection of simulated parameter values needs to be
calibrated. We propose to conduct likelihood-free Bayesian inferences about parameters with
no prior selection of the relevant components of the summary statistics and
bypassing the derivation of the associated tolerance level. The approach relies on the
random forest methodology of \citet{breiman:2001} applied in a (non parametric) regression setting.
We advocate the derivation of a new random forest for each component of
the parameter vector of interest. When compared with earlier ABC solutions, this method
offers significant gains in terms of robustness to the choice of the summary
statistics, does not depend on any type of tolerance level, and is a good trade-off in term of quality of point estimator precision and credible interval estimations for a given computing time. We illustrate the performance of our methodological proposal and compare it with earlier ABC methods on a Normal toy example and a population genetics example dealing with human population evolution. All methods designed here have been incorporated in the {\sf R} package \texttt{abcrf} (version 1.7) available on CRAN.

\vspace{0.5cm} \noindent \textbf{Keywords:} 
Approximate Bayesian computation, Bayesian inference, likelihood-free methods, parameter inference, random forests
\end{abstract}

\newpage

\section{Introduction}

As statistical models and data structures get increasingly complex, managing
the likelihood function becomes a more and more frequent issue. We now face
many realistic fully parametric situations where the likelihood function cannot
be computed in a reasonable time or simply is unavailable. As a result, while
the corresponding parametric model is well-defined, with unknown parameter
$\theta$, standard solutions based on the density function $f(y\mid \theta)$
like Bayesian or maximum likelihood analyses are prohibitive to implement. To
bypass this hurdle, the last decade witnessed different inferential strategies,
among which composite likelihoods \citep{lindsay:1988,varin:reid:firth:2011},
indirect inference \citep{gourieroux:monfort:renault:1993}, GMMs
\citep{chernozhukov:han:2003}, and likelihood-free methods such as 
approximate Bayesian computation \citep[ABC,][]{beaumont:zhang:balding:2002,csillery:blum:gaggiotti:francois:2010,
marin:pudlo:robert:ryder:2012}, became popular options. We focus here on
improving the latter solution.

Since their introduction in population genetics
\citep{tavare:balding:griffith:donnelly:1997,
pritchard:seielstad:perez:feldman:1999,beaumont:zhang:balding:2002}, ABC
methods have been used in an ever increasing range of applications,
corresponding to different types of complex models in diverse scientific
fields \citep[see, e.g.,][]{beaumont:2008, toni:etal:2009, beaumont:2010,
csillery:blum:gaggiotti:francois:2010, theunert:etal:2012, chan:etal:2014,
arenas:etal:2015, sisson:etal:2017}.

Posterior distributions are the cornerstone of any Bayesian analysis as they
constitute both a sufficient summary of the data and a means to deliver all
aspects of inference, from point estimators to predictions and uncertainty
quantification. However, it is rather common that practitioners and users of
Bayesian inference are not directly interested in the posterior distribution {\em per
se}, but rather in some summary aspects, like posterior mean, posterior
variance or posterior quantiles, since these are easier to interpret and report.
With this motivation, we consider a version of ABC focussing on the
approximation of unidimensional transforms of interest like the above, instead
of resorting to the classical ABC approach that aims at approximating the
entire posterior distribution and then handling it as in regular Bayesian
inference. The approach we study here is based on random forests
\citep[RF,][]{breiman:2001}, which produces non-parametric regressions on an
arbitrary set of potential regressors. We recall that the calibration side of RF (i.e. the choice of the RF parameters: typically the number of trees and the number of summary statistics sampled at each node) was successfully exploited in \citet{pudlo:etal:2016} for conducting ABC
model choice.

After exposing the ABC and RF principles, we explain how to fuse both
methodologies towards Bayesian inference about parameters of interest. We then
illustrate the performance of our proposal and compare it with earlier ABC methods on a Normal toy example and a
population genetics example dealing with human population evolution.


\section{Methods}
\label{sec:methodology}

Let $$\{f(y \mid \theta)\colon y \in \mathcal{Y}, \theta \in \Theta\}\,,\
\mathcal{Y} \subseteq \mathbb{R}^n\,,\ \Theta \subseteq \mathbb{R}^p\,\qquad p,
n \geq 1$$ be a parametric statistical model and $\pi(\theta)$ be a prior
distribution on the parameter $\theta$.  Given an observation (or sample) $y$
issued from this model, Bayesian parameter inference is based on the posterior
distribution $\pi(\theta \mid y)\propto \pi(\theta)f(y|\theta)$. The
computational difficulty addressed by ABC techniques is that a numerical
evaluation of the density (a.k.a., likelihood) $f(y\mid \theta)$ is impossible
or at least very costly, hence preventing the derivation of the posterior
$\pi(\theta \mid y)$, even by techniques like MCMC \citep{marin:robert:2014}.

\subsection{ABC for parameter inference}
\label{subsec:rappelABC}

The principle at the core of ABC is to approximate traditional Bayesian
inference from a given dataset by simulations from the prior distribution. The
simulated values are accepted or rejected according to the degree of
proximity between the observed dataset $y$ and a simulated one $y(\theta)$
thanks to a (usually normalized Euclidean) distance $d$. ABC relies on the
operational assumptions that, while the likelihood is intractable, observations can be
generated from the statistical model $f(\cdot\mid \theta)$ under consideration
for a given value of the parameter $\theta$.

The ABC resolution of the intractability issue with the likelihood is to produce a
so-called \textit{reference table}, recording a large number of datasets simulated from the prior
predictive distribution, with density $f(y \mid \theta)\times\pi(\theta)$, and then
extracting those that bring the simulations close enough to the actual sample.
In most ABC implementations, for both computational and statistical efficiency
reasons, the simulated $y^{(t)}$'s $(t=1,\ldots,N)$ are summarised through a
dimension-reduction function $\eta\colon \mathcal{Y} \to \mathbb{R}^k$, often called
a vector of $k$ summary statistics. While the outcome of the ABC
algorithm is then an approximation to the posterior distribution of $\theta$
given $\eta(y)$, rather than given the entire data $y$
\citep{marin:pudlo:robert:ryder:2012}, arguments are to be found in the
literature supporting the (ideal) choice of a summary statistic $\eta$ of the
same dimension as the parameter \citep{fearnhead:prangle:2012,li:fearnhead:2015,frazier:etal:2017}.
Algorithm~\ref{alg:reference-table} details how the \textit{reference table} is
constructed. The \textit{reference table} will latter be used as a training
dataset for the different RF methods explained below.

In practice, a \textit{reference table} of size $N$ is simulated,
distances $\big( d(\eta(y),\eta(y^{(t)})) \big)_{t=1,\ldots,N}$ are computed
and then given a tolerance proportion $0<p_\epsilon \leq 1$, pairs $(\theta^{(t)},
\eta(y^{(t)}))$ within the $p_\epsilon$ range of lowest distances are selected.
The parameter sample thus derived is deemed to approximate the
posterior distribution $\pi( \theta \mid \eta(y) )$.

\begin{algorithm}
  \For{$t \leftarrow 1$ \KwTo $N$}{
    Simulate $\theta^{(t)} \sim \pi(\theta)$\;
    Simulate $y^{(t)} = (y_1^{(t)}, \ldots, y_n^{(t)}) \sim f\left(y \mid \theta^{(t)}\right)$\;
    Compute $\eta(y^{(t)}) = \{\eta_1(y^{(t)}), \ldots, \eta_k(y^{(t)})\}$\;
  }
  \caption{Generation of a \textit{reference table} from the prior predictive distribution ${\pi(\theta) f(y \mid \theta)}$}
  \label{alg:reference-table}
\end{algorithm}

The method is asymptotically consistent in the sense that the true parameter
behind the data can be exhibited when both the sample size and the number of
simulations grow to infinity and the tolerance decreases to zero
\citep{frazier:etal:2017}. However, it suffers from two major drawbacks. First,
to ensure a sufficient degree of reliability, the number $N$ of simulations
must be quite large. Hence, it may prove difficult to apply ABC on large or
complex datasets since producing data may prove extremely costly. Second, the
calibration of the ABC algorithm (i.e. a tolerance level indicating the separation
of accepted from rejected simulated parameter values) is a critical step and impacts the resulting approximation
\citep{marin:pudlo:robert:ryder:2012,blum:nunes:prangle:sisson:2013}. Since
the justification of the method is doubly asymptotic, it is delicate if not impossible 
\citep{li:fearnhead:2015,frazier:etal:2017} to optimally tune ABC for finite
sample sizes. A third feature of major importance in this algorithm is
that it requires selecting a vector of summary statistics that captures enough
information from the observed and simulated data. For most problems, using
the raw data to run the comparison is indeed impossible, if only because of the
dimension of the data. \cite{fearnhead:prangle:2012} give a
natural interpretation of the vector of summary statistics as an estimator
of $\theta$, but this puts a clear restriction on the dimension and nature of
the components of $\eta(y)$.

Finally, it has to be noted that the original rejection ABC method, which can
be interpreted as a $K$-nearest neighbour method, has been recurrently improved by
linear or non-linear regression strategies, mentioned in literature as adjusted
local linear \citep{beaumont:zhang:balding:2002}, ridge regression
\citep{blum:nunes:prangle:sisson:2013}, and by methods based on adjusted neural networks
\citep{blum:francois:2010}. Instead of ridge regularization within local linear adjustment, 
\cite{saulnier:etal:2017} propose to use a lasso regularization in order to select among the summary
statistics. The obtained results are promising exceptt when the summary statistics are
highly correlated. In such cases, \cite{saulnier:etal:2017} suggest to use random forests.We will only consider ridge regularization in the present work.

\subsection{Random forest methodology}

The random forest methodology (RF) of \cite{breiman:2001} is pivotal in our
proposal. We use Brei\-man's RF in a regression setting where a response variable
$Y \in \mathbb{R}$ is explained by a vector of covariates $X=(X^{(1)},
\ldots, X^{(k)})$. A collection of $N$ datasets, made of responses and associated covariates, is used to train a RF.

A given regression RF of size $B$ is composed of $B$ regression
trees. A tree is a structure made of binary nodes, which are iteratively built from top to
bottom  until a stopping rule is satisfied. There are two types of nodes in such trees, the internal
and terminal nodes, the latter being also called leaves. At an internal node, a binary
rule of the form $X^{(j)} \leq s$ versus $X^{(j)} > s$ compares a covariate $X^{(j)}$ with a bound $s$.
The result of the test divides the predictor space and the training dataset
depending on this splitting rule into two parts in two new different nodes.
When constructing the tree based on a training sample, the covariate index $j$ and the splitting bound $s$
are determined towards minimising a $L^2$-loss criterion. The same covariate may be used multiple times for the choice of $j$ at different levels of the tree construction. 
Splitting events stop when all the observations of the training dataset in a given node have the same covariates value, in that case the node becomes a leaf. 
Moreover, when a node has less than $N_\text{min}$ observations, the node also becomes a leaf, typically $N_\text{min}=5$ in the regression framework.
Once the tree construction is complete, a value of the response variable is allocated to each tree leaf, corresponding to the average of the response values of the present datasets.
For a given and an observed dataset that corresponds to a new covariate X, predicting the associated value of Y implies following the path of the binary rules. The outcome of the prediction is the allocated value of the leaf where this dataset ends after following this path.

The RF method consists in aggregating (or bagging) randomized
regression trees. A large number of trees are trained on bootstrap samples of
the training dataset and
furthermore a subset of $n_\text{try}$ covariates among the $k$ available covariates are randomly
considered at each split. The predicted value of a regression RF is determined by
averaging the $B$ predictions over its $B$ tree components.

\subsection{ABC parameter inference using random forest}

\subsubsection{Motivations and main principles}

The particular choice of RF as a (non-parametric) estimation method in a regression setting is justified by
the robustness of both random forests and quantile methods to ``noise", that is, to
the presence of irrelevant predictors, even when the proportion of such covariates
amongst the entire set of proposed predictors is substantial \citep{marin:etal:2017}.  
By comparison, the method of $K$-nearest neighbour classifiers lacks such
characteristics \citep{biau:etal:2015}. In the setting of building an ABC
algorithm without preliminary selection of some summary statistics, our
conjecture is that RF allows for the inclusion of an arbitrary and potentially
large number of summary statistics in the derivation of the forest and therefore
that it does not require the usual preliminary selection of summary statistics.
When implementing this approach, we hence bypass the selection of summary statistics and
include a large collection of summary statistics, some or many of which being potentially poorly informative if not irrelevant for the regression.
For earlier considerations on the selection of summary statistics, see
\citet{joyce:marjoram:2008,nunes:balding:2010,
jung:marjoram:2011,fearnhead:prangle:2012} and the review paper of
\citet{blum:nunes:prangle:sisson:2013} where different dimension reduction
techniques are compared.

A regression RF produces an expected predicted value for an arbitrary
transform of $\theta$, conditional on an observed dataset. This prediction is
the output of a piece-wise constant function of the summary
statistics.
RF aggregates trees, partitions the feature space (here the space
of summary statistics) in a way tuned to the forecast of a scalar output, i.e., a
one dimensional functional of the parameter. This partition and prediction are done without requiring the definition of a particular distance on the feature space and is hence not dependant of any type of tolerance level. From an ABC perspective, each tree
of a RF provides a partition of the summary space that is
adapted for the forecasting of a scalar transform $h(\theta)$ of the
parameter $\theta$. In the following subsection we present how to compute quantities of interest in a context of parameter inference, thanks to the calculation of weights.

\subsubsection{Calculation of weights and approximation of the posterior expectation}

Assume we have now grown a RF made of $B$ trees that predicts
${\tau=h(\theta) \in \mathbb{R}}$ using $\eta(y)$ and the training sample
$(\eta(y^{(t)}), \tau^{(t)})_{t=1,\ldots,N}$,
where $\tau^{(t)}=h(\theta^{(t)})$. In the examples below, we will consider the case where $h$ is the projection on a given coordinate of $\theta$. Each of these $B$ trees produces a
partition of the space of summary statistics, with a constant prediction of
the expected value of $\tau$ on each set of the partition. More precisely, given
$b$-th tree in the forest, let us denote $n_b^{(t)}$ the number of times the pair
$(\eta(y^{(t)}), \tau^{(t)})$ is repeated in the bootstrap sample that
is used for building the $b$-th tree. Note that $n_b^{(t)}$ is equal to zero when the
pair does not belong to the bootstrap sample. These pairs form the so-called
out-of-bag sample of the $b$-th tree. Now, let $L_b(\eta(y))$
denote the leaf reached after following the path of binary choices
given by the tree, which depends on the value of $\eta(y)$. 
The number of items of the bootstrap sample that fall in that leaf is
\[
\big|L_b(\eta(y))\big| = \sum_{t=1}^N n_b^{(t)} \mathbf
1\Big\{\eta(y^{(t)})\in L_b(\eta(y))\Big\}\,,
\]
where $\mathbf{1}$ denotes the indicator function,
and the mean value of $\tau$ of that leaf of the $b$-th tree is
\[
\frac{1}{\big|L_b(\eta(y))\big|}\sum_{t=1}^N n_b^{(t)} \mathbf
1\Big\{\eta(y^{(t)})\in L_b(\eta(y))\Big\} \tau^{(t)}.
\]
Averaging these $B$ predictions of $\tau$ leads to an approximation of
the posterior expected value of $\tau$, also denoted mean value of $\tau$, which can be written as follows:
\[
\widetilde{\mathbb{E}}\big(\tau\big|\eta(y)\big) = \frac{1}{B}\sum_{t=1}^N
\sum_{b=1}^B \frac{1}{\big|L_b(\eta(y))\big|} n_b^{(t)} \mathbf
1\Big\{\eta(y^{(t)})\in L_b(\eta(y))\Big\} \tau^{(t)}.
\]
As exhibited by \cite{meinshausen:2006}, the above can be seen as a weighted
average of $\tau$ along the whole training sample of size $N$ made by
the \textit{reference table}. In fact, the weight of the $t$-th pair
$(\eta(y^{(t)}), \tau^{(t)})$ given $\eta(y)$ is
\[
w_t(\eta(y)) = \frac{1}{B}\sum_{b=1}^B \frac{1}{\big|L_b(\eta(y))\big|} n_b^{(t)} \mathbf
1\Big\{\eta(y^{(t)})\in L_b(\eta(y))\Big\}.
\]

\subsubsection{Approximation of the posterior quantile and variance}

The weights $w_t(\eta(y))$ provide an approximation of the
posterior cumulative distribution function (cdf) of $\tau$ given $\eta(y)$ as
\begin{equation*}
\widetilde{F}(\tau \mid \eta(y) ) = \sum_{t=1}^N w_t(\eta(y)) \mathbf
1{\{ \tau^{(t)} < \tau \} }.
\end{equation*}
Posterior quantiles, and hence credible intervals, are then derived by inverting this empirical cdf, that is by plugging
$\tilde{F}$ in the regular quantile definition
\begin{align*}
\widetilde{\mathbb{Q}}_\alpha \{ \tau \mid \eta(y) \} = \mbox{inf}\left\{\tau : \widetilde{F}(\tau \mid \eta(y) ) \geq
\alpha \right\}.
\end{align*}
This derivation of a quantile approximation is implemented in the {\sf R} package \texttt{quantregForest} and 
the consistency of $\tilde{F}$ is established in \citet{meinshausen:2006}. 

An approximation of $\mbox{Var}(\tau \mid y)$ can be derived in a natural way from $\widetilde{F}$, leading to 
\begin{equation*}
\widehat{\mbox{Var}}(\tau \mid \eta(y) ) = \sum_{t=1}^N w_t(\eta(y)) \left( \tau^{(t)} -  \sum_{u=1}^N w_u(\eta(y)) \tau^{(u)} \right)^2.
\end{equation*}

\subsubsection{Alternative variance approximation}
\label{subsec:altvar}

Regarding the specific case of the posterior variance of $\tau$, we propose a
slightly more involved albeit manageable version of a variance estimate. 
Recall that, in any given tree $b$, some entries from the \textit{reference table} are not included since each tree relies on a bootstrap sample of the training dataset. The out-of-bag simulations,
i.e. unused in a bootstrap sample, can be exploited toward returning
an approximation of $\mathbb{E}\{\tau \mid \eta(y^{(t)})\}$,
denoted $\hat\tau_{ \text{oob} }^{(t)}$.  Indeed, given a vector of summary statistics $\eta(y^{(t)})$ of the training dataset, passing this vector down the ensemble of trees where it has not been used and mean averaging the associated predictions provide such an approximation.
Since
\begin{equation*}
\mbox{Var}(\tau \mid \eta(y)) = \mathbb{E} \left(\left[\tau 
- \mathbb{E}\{\tau \mid \eta(y)\} \right]^2 \mid \eta(y) \right)\,,
\end{equation*}
we advocate applying the original RF weights $w_t(\eta(y))$ to the out-of-bag square residuals
$(\tau^{(t)} - \hat{\tau}_{ \text{oob} }^{(t)})^2$, which results in the alternative approximation
\begin{equation*}
\widetilde{ \mbox{Var} } (\tau \mid \eta(y)) = \sum_{t=1}^N w_t\{\eta(y)\}
(\tau^{(t)} - \hat{\tau}_{ \text{oob} }^{(t)})^2.
\end{equation*}
Under the same hypotheses as \cite{meinshausen:2006}, this estimator converges when $N\rightarrow\infty$. Indeed, $\hat{\tau}_{ \text{oob} }^{(t)}$ and $\sum_{t=1}^N w_t(\eta(y)) \tau^{(t)}$ tends to the same posterior expectation. Hence, the two variance estimators above mentioned are equivalent.
A comparison between different variance estimators is detailed in the supplementary
material. Owing to the results of this comparative study, we choose to use the above alternative variance estimator when presenting the results from two examples.

As a final remark, it is worth stressing that the approximation of the posterior covariance 
between a pair of parameters can be achieved thanks to a total of three RFs. The details of that statistical extension are presented in the Section \ref{sup:sec:cov} of the supplementary material.

\subsubsection{A new {\sf R} package for conducting parameter inferences using ABC-RF}

When several parameters are jointly of interest, our recommended global strategy consists
in constructing one independent RF for each parameter of interest and estimate from each RF several summary measurements of the posterior distribution (i.e. posterior expectation, quantiles and variance) of each parameter. Additional RFs might be constructed, however, if one is interested by estimating posterior covariance between pair of parameters.

Our {\sf R} library \texttt{abcrf} was initially developed for Bayesian model choice using
ABC-RF as in \citet{pudlo:etal:2016}. The version 1.7 of \texttt{abcrf} includes
all the methods proposed in this paper to estimate posterior expectations, quantiles, variances
(and covariances) of parameter(s). \texttt{abcrf} version 1.7 is available on
CRAN. We provide in the Section \ref{sup:sec:codeR} of the supplementary material, a commented {\sf R} code that will allow non expert users to run random forest inferences about parameters using the \texttt{abcrf} package version 1.7.

\section{Results from two examples}
\label{sec:simulation}

We illustrate the performances of our ABC-RF method for Bayesian parameter
inference on a Normal toy-example and on a realistic population genetics example. In
the first case and only in that case, approximations of posterior quantities can be
compared with their exact counterpart. For both examples, we further compare the
performances of our methodology with those of earlier ABC methods based on
solely rejection, adjusted local linear \citep{beaumont:zhang:balding:2002},
ridge regression \citep{blum:nunes:prangle:sisson:2013}, and adjusted neural
networks \citep{blum:francois:2010}.

For both illustrations, RFs were trained based on the functions of the {\sf R}
package \texttt{ranger} \citep{wright:ziegler:2017} with forests made of $B=500$
trees, with $n_\text{try}=k/3$ selected covariates (i.e. summary statistics) 
for split-point selection at each node, and with a minimum node size equals to $5$ \citep[][and see Section \ref{subsec:practical} Practical recommendations regarding the implementation of the ABC-RF algorithm]{breiman:2001}. The other ABC methods in the comparison were based on the same \textit{reference
tables}, calling the corresponding functions in the {\sf R} package \texttt{abc}
\citep{csillery:francois:blum:2012, nunes:prangle:2015} with
its default parameters. ABC with neural network adjustment require
the specification of the number of layers composing the neural network: we opted for
the default number of layers in the {\sf R} package \texttt{abc}, namely $10$.
A correction for heteroscedasticity is applied by default when considering regression
adjustment approaches. Note that regression corrections are univariate for local linear and ridge regression as well as for RF, whereas neural network - by construction - performs multivariate corrections.

\subsection{Normal toy example}
\label{subsec:gaussian}

We consider the hierarchical Normal mean model
\begin{align*}
  y_i \mid \theta_1, \theta_2 &\sim \mathcal{N}(\theta_1, \theta_2),\\
  \theta_1 \mid \theta_2 &\sim \mathcal{N}(0, \theta_2), \\
  \theta_2 &\sim I\mathcal{G}(4,3),
\end{align*}
where $I\mathcal{G}(\kappa, \lambda)$ denotes an inverse Gamma distribution
with shape parameter $\kappa$ and scale parameter $\lambda$. Let $y = (y_1,
\ldots, y_n)$ be a $n$-sample from the above model. Given these conjugate
distributions, the marginal posterior distributions
of the parameters $\theta_1$ and $\theta_2$ are closed-forms:
\begin{align*}
\theta_1 \mid y &\sim \mathcal{T} \left(n+8, \frac{n\bar y}{n + 1}, \frac{2 \left( 3 + s^2/2 + n {\bar y}^2 / (2n+2) \right)}{(n + 1)(n + 8)} \right) \\
\theta_2 \mid y &\sim I\mathcal{G}\left( \frac{n}{2}+4, 3 + \frac{s^2}{2} +  \frac{n {\bar y}^2}{2n+2} \right),
\end{align*}
where $\bar y$ is the sample mean and $s^2=\sum_{i=1}^n(y_i - \bar y)^2$ the sum of squared deviations.
$\mathcal{T}(\nu, a, b)$ denotes the general $t$ distribution
with $\nu$ degrees of freedom \citep{marin:robert:2014}.

From the above expressions and for a given sample $y$, it is straightforward to derive the exact values of
$\mathbb{E}(\theta_1 \mid y)$, $\mbox{Var}(\theta_1 \mid y)$, $\mathbb{E}(\theta_2 \mid y)$,
$\mbox{Var}(\theta_2 \mid y)$ and posterior quantiles for the two parameters. This provides us with a benchmark on which to assess the performances of ABC-RF.
For the present simulation study, we opted for a \textit{reference table} made of $N = 10^4$ replicates of a sample of size $n = 10$ and
$k = 61$ summary statistics. Those statistics included the sample mean, the sample variance, the sample median absolute
deviation (MAD), all possible sums and products with these three elements resulting in eight new summary statistics
and $50$ additional independent (pure) noise variables that were generated from a 
uniform $\mathcal{U}_{[0,1]}$ distribution. 
The performances of our method were evaluated on an
independent test table of size $N_{\rm pred} = 100$, produced in the same way
as the \textit{reference table}. Current ABC methods (rejection, adjusted
local linear, ridge and neural network) all depend on the choice of a tolerance level $p_\epsilon$ corresponding to the proportion of
selected simulated parameters with lowest distances between
simulated and observed summary statistics. On this example we
consider a tolerance level of $p_\epsilon=0.01$ for ABC with rejection, and $p_\epsilon=0.1$
for the ABC methods with adjustment.

\begin{figure}
  \centering
	\includegraphics[width=0.49\linewidth]{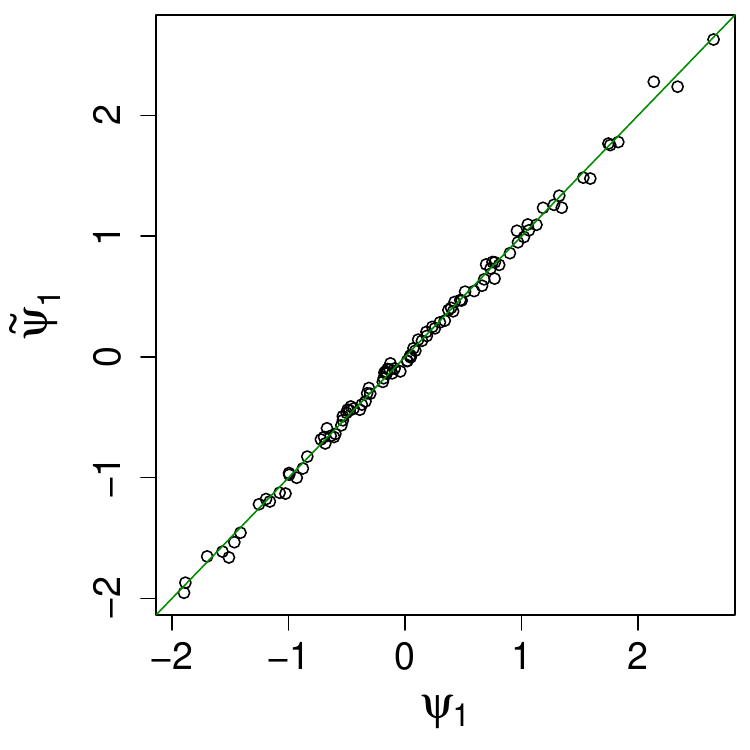}
	\includegraphics[width=0.49\linewidth]{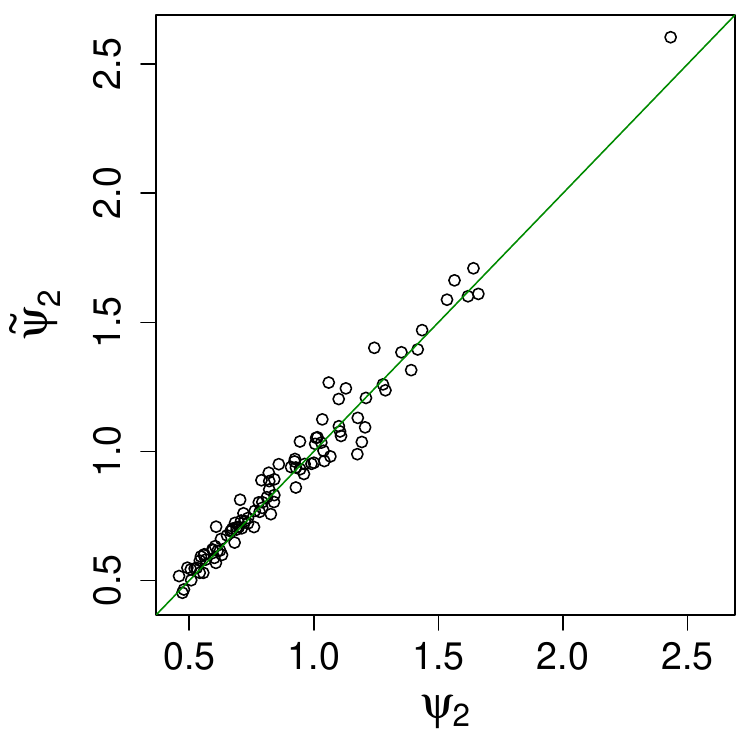}
	\includegraphics[width=0.49\linewidth]{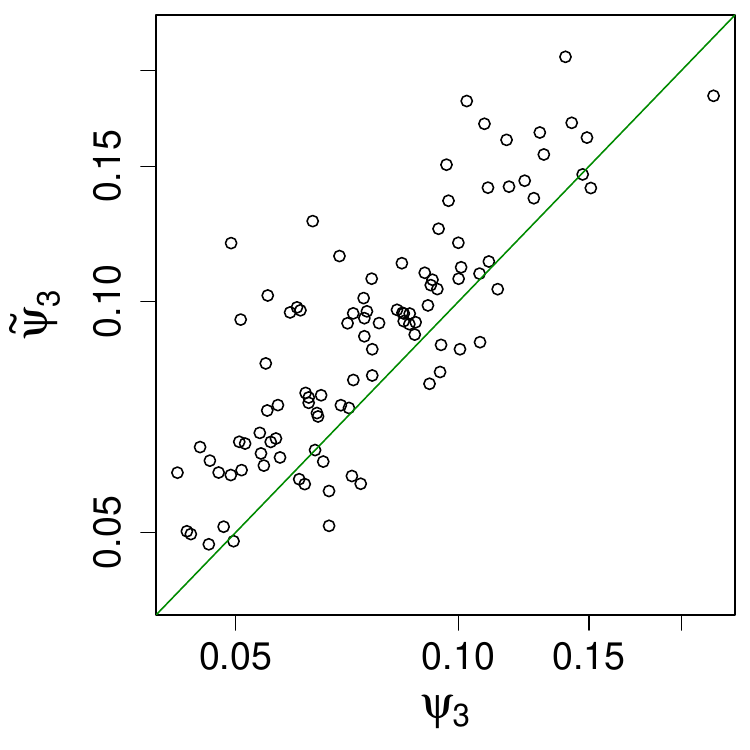}
	\includegraphics[width=0.49\linewidth]{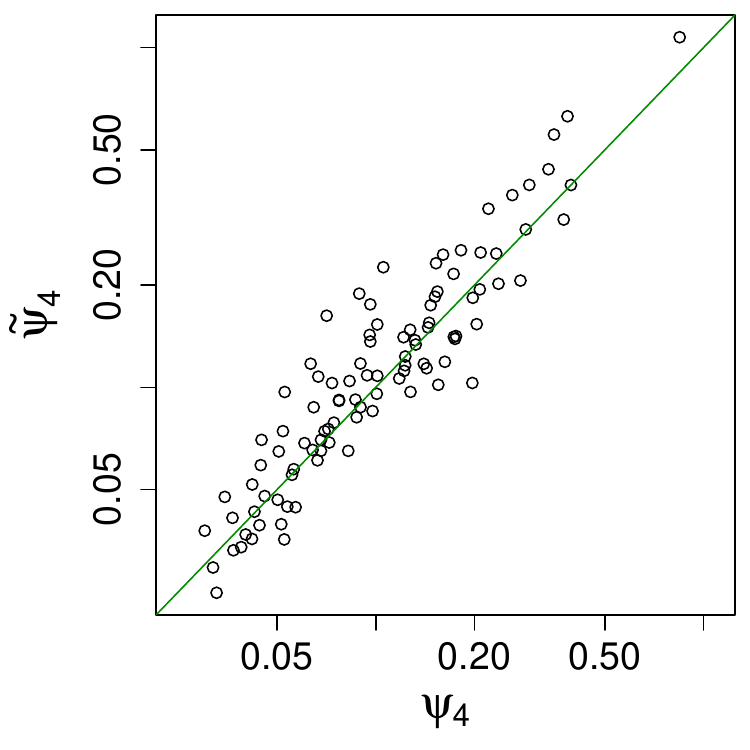}
  \caption{Scatterplot of the theoretical values
    $\psi_1(y) = \mathbb{E}(\theta_1 \mid y)$, $\psi_2(y) = \mathbb{E}(\theta_2 \mid y)$, ${\psi_3(y) = \mbox{Var}(\theta_1 \mid y)}$ and $\psi_4(y) = \mbox{Var}(\theta_2 \mid y)$ for the Normal model with
    their corresponding estimates $\tilde{\psi}_1$, $\tilde{\psi}_2$, $\tilde{\psi}_3$, $\tilde{\psi}_4$
    obtained using ABC-RF. Variances are represented on a log scale.}
  \label{fig:QQplot-expectation-variance}
\end{figure}

\begin{figure}
  \centering
	\includegraphics[width=0.49\linewidth]{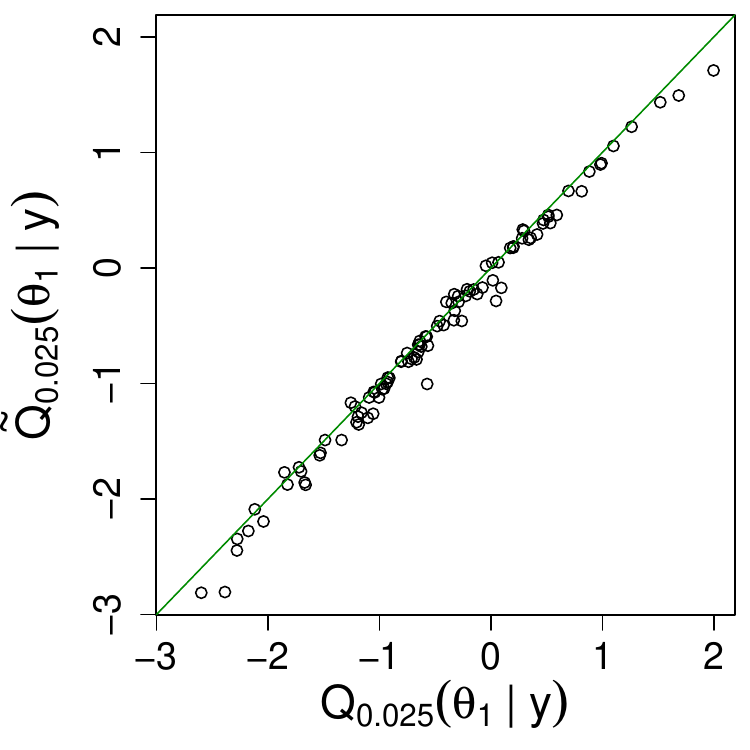}
	\includegraphics[width=0.49\linewidth]{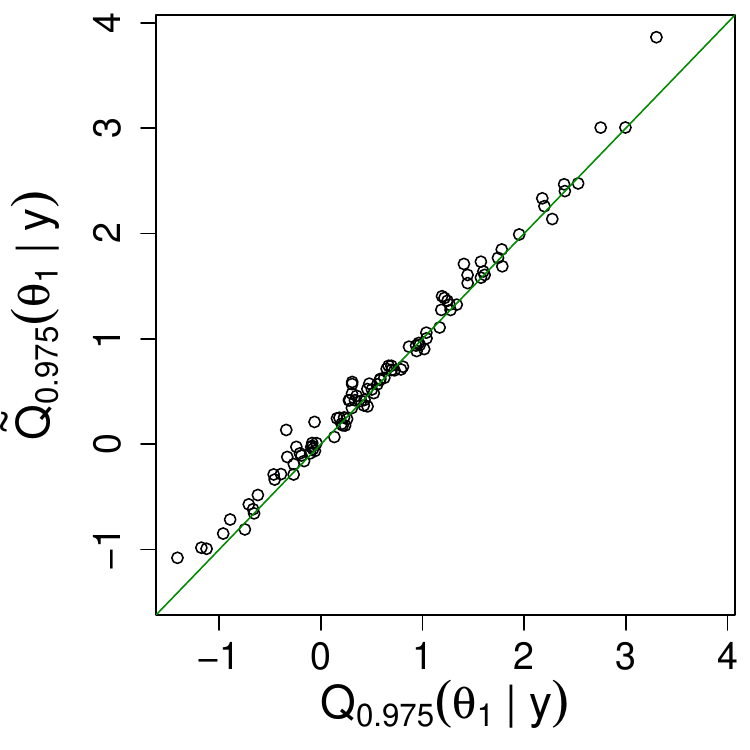}
	\includegraphics[width=0.49\linewidth]{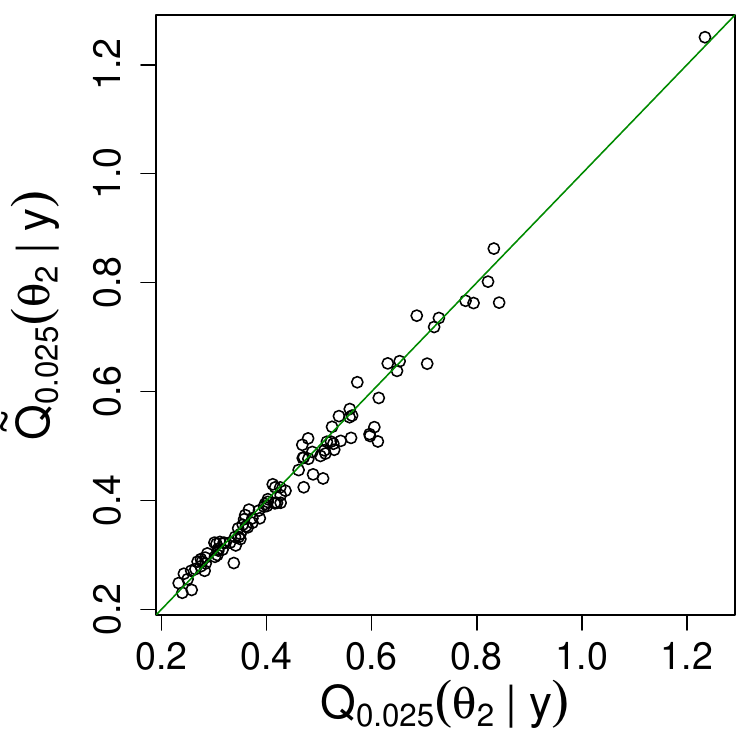}
	\includegraphics[width=0.49\linewidth]{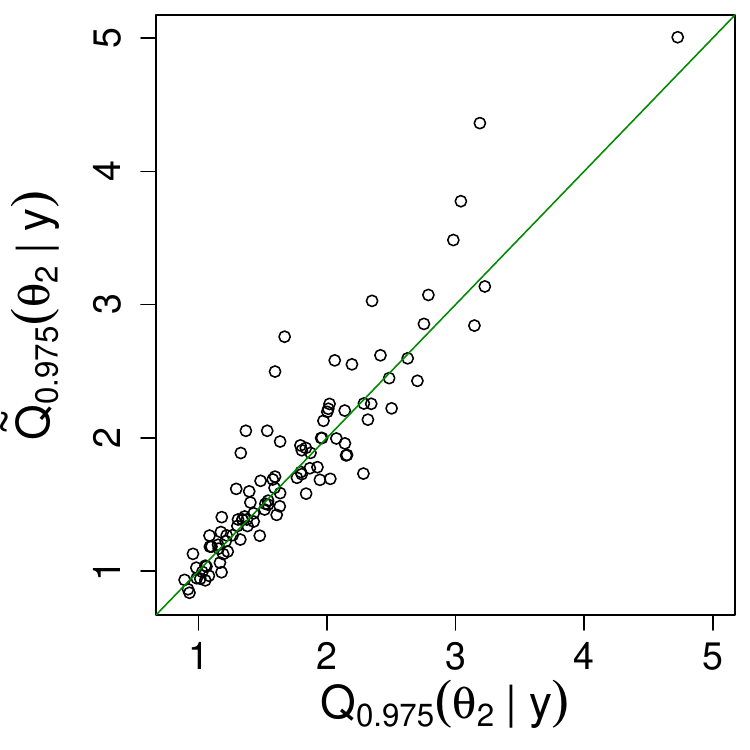}
  \caption{Scatterplot of the theoretical values of $2.5\%$ and $97.5\%$ posterior quantiles for $\theta_1$ and $\theta_2$,
  for the Normal model with their corresponding estimates obtained using ABC-RF.}
  \label{fig:QQplot-quantiles}
\end{figure}

Figure~\ref{fig:QQplot-expectation-variance} compares the exact values $\psi_1(y)
= \mathbb{E}(\theta_1 \mid y)$, $\psi_2(y) = \mathbb{E}(\theta_2 \mid y)$, $\psi_3(y)
= \mbox{Var}(\theta_1 \mid y)$ and $\psi_4(y) = \mbox{Var}(\theta_2 \mid y)$ with
the estimates obtained from the ABC-RF approach. It shows that the proposed
estimators have good overall performances for both $\psi_1(y)$ and $\psi_2(y)$,
although one can see that $\psi_2(y)$ tends to be slightly overestimated. Our
estimators perform less satisfactorily for both $\psi_3(y)$ and
$\psi_4(y)$ but remain acceptable. Figure~\ref{fig:QQplot-quantiles} shows furthermore that the
quantile estimation are good for $\theta_1$ if less accurate for $\theta_2$.

We then run an experiment to evaluate the precision of the marginal posterior approximation provided by ABC-RF for the parameter $\theta_1$, using two different test datasets and 40 independent \textit{reference tables}. As exhibited in Figure \ref{fig:true_vs_RFcdfs}, results are mixed. 
For one dataset, the fit is quite satisfactory, with the RF approximation
showing only slightly fatter tails than the true posterior density distribution function (Figure
\ref{fig:true_vs_RFcdfs}; upper panel). For the other dataset, we obtain
stronger divergence both in location and precision of the posterior density distribution function
(Figure \ref{fig:true_vs_RFcdfs}; lower panel).

\begin{table}
\centerline{
\begin{tabular}{c c c c c c c}
 & RF & Reject & ALL & ARR & ANN \\
\hline $\psi_1(y) = \mathbb{E}(\theta_1 \mid y)$ 	& \textbf{0.18} & 0.32 & 0.34 & 0.31 & 0.42 \\
$\psi_2(y) = \mathbb{E}(\theta_2 \mid y)$ 			& \textbf{0.05} & 0.10 & 0.14 & 0.17 & 0.17 \\
$\psi_3(y) = \hbox{Var}(\theta_1 \mid y)$ 			& \textbf{0.25} & 2.21 & 0.70 & 0.69 & 0.48 \\
$\psi_4(y) = \hbox{Var}(\theta_2 \mid y)$ 			& \textbf{0.25} & 0.43 & 0.66 & 0.70 & 0.97 \\
$Q_{0.025}(\theta_1|y)$ 							& \textbf{0.34} & 1.61 & 0.69 & 0.84 & 0.50 \\
$Q_{0.025}(\theta_2|y)$ 							& \textbf{0.04} & 0.13 & 0.34 & 0.55 & 0.80 \\
$Q_{0.975}(\theta_1|y)$ 							& \textbf{0.25} & 1.35 & 0.53 & 0.70 & 0.60 \\
$Q_{0.975}(\theta_2|y)$ 							& \textbf{0.10} & 0.14 & 0.20 & 0.20 & 0.42 \\
\end{tabular}
}
\caption{\label{tab:gaus-NMAE} Comparison of normalized mean absolute errors
(NMAE) of estimated quantities of interest obtained with ABC-RF and other ABC methodologies. RF, Reject, ALL, ARR and ANN stand for random forest (ABC-RF), rejection, adjusted local linear, adjusted ridge regression and adjusted neural network methods, respectively. The smallest NMAE values are in by bold characters.}
\end{table}

\begin{figure}
  \centering
  \includegraphics[width=0.49\linewidth]{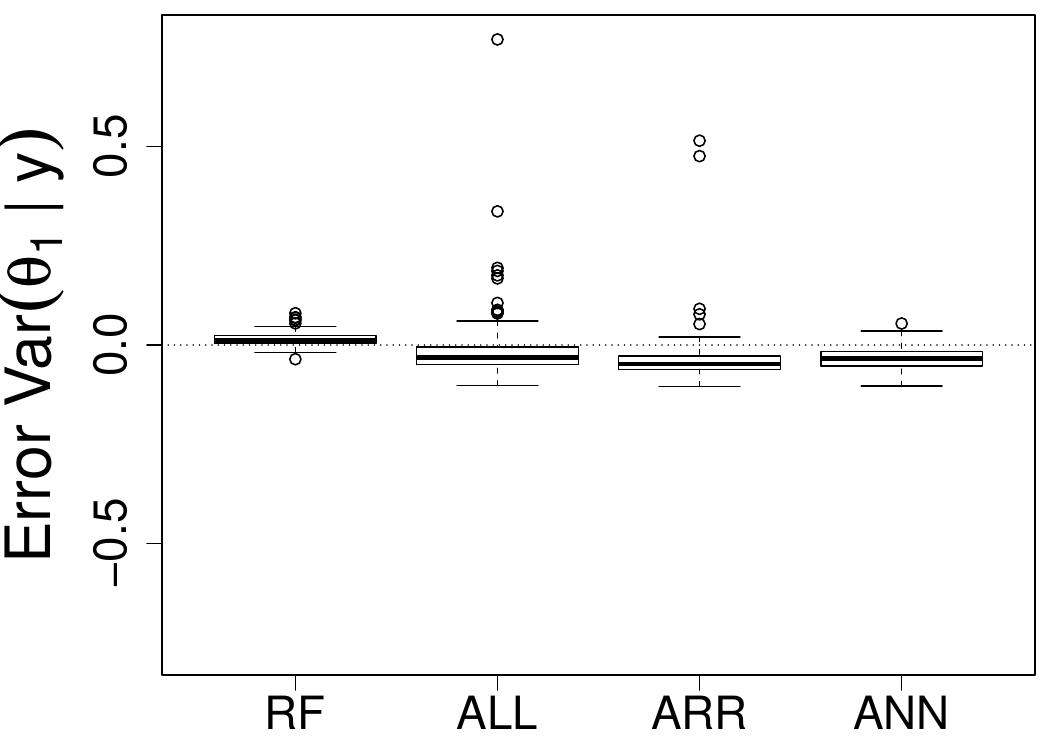}
  \includegraphics[width=0.49\linewidth]{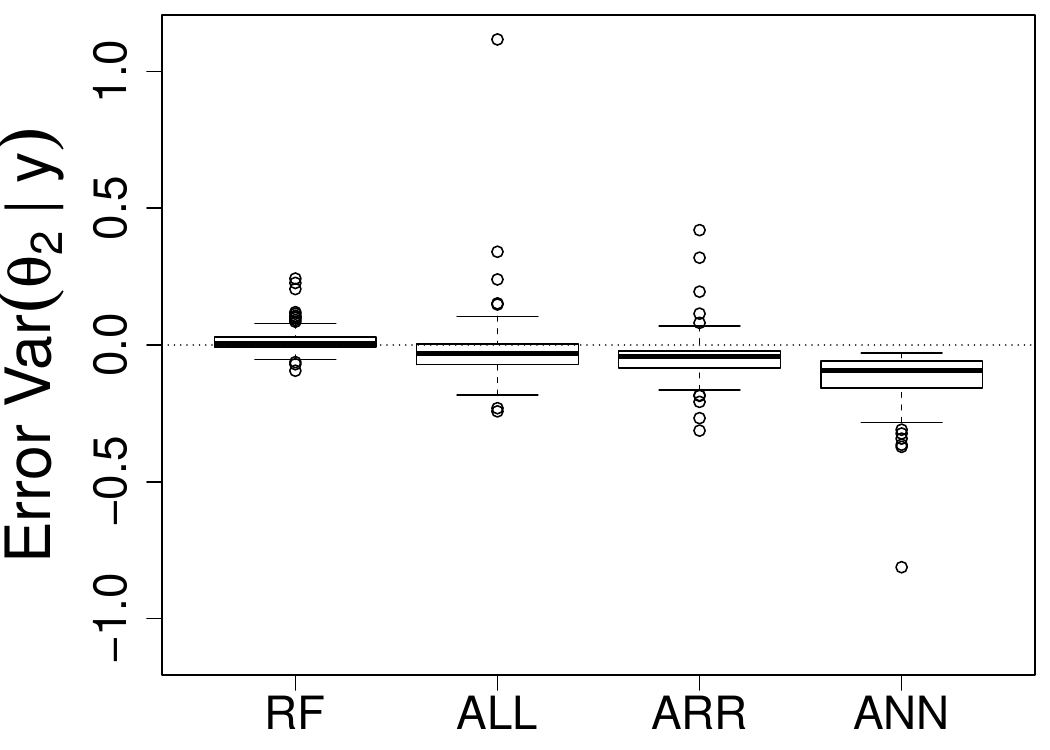}
  \caption{Boxplot comparison of the differences between our predictions for
$\mbox{Var}(\theta_1 \mid y)$ and $\mbox{Var}(\theta_2 \mid y)$ and the corresponding true values, using ABC-RF and other ABC methods. RF, ALL, ARR and ANN notations as in the legend of Table \ref{tab:gaus-NMAE}). The closer to the $y=0$ axis, the better the predictions. Boxplots above this axis imply overestimation of the predictions and below underestimation. }
  \label{fig:normal-boxplot-error-variances}
\end{figure}

Using the same \textit{reference table}, we now compare our ABC-RF results with
a set of four earlier ABC methods, namely, ABC methods based on straightforward
rejection, adjusted local linear, ridge
regression and adjusted neural networks.
Normalized mean absolute errors are used to measure performances. 
The normalization being done by dividing the absolute error by the true value of the target.
A normalized version offers the advantage of being hardly impacted when
only a few observations get poorly predicted. Table~\ref{tab:gaus-NMAE} shows
that the ABC-RF approach leads to results better than all other ABC methods for
all quantities of interest. Expectations and quantiles are noticeably more
accurately predicted. Figure~\ref{fig:normal-boxplot-error-variances} compares
differences between estimated and true values of the posterior variances
$\psi_3(y)$, $\psi_4(y)$.  It shows the global underestimation associated with
earlier ABC methods, when compared to ABC-RF, the latter only slightly
overestimating the posterior variance. Finally, by looking at the width of
the boxplots of Figure \ref{fig:normal-boxplot-error-variances}, we deduce that
our ABC-RF estimations exhibits a lower estimation variability.

\subsection{Human population genetics example}
\label{subsec:human}

\begin{figure}
  \centering
  \adjincludegraphics[width=0.8\linewidth, trim={{.25\width} {.2\width} {.25\width} {.1\width}},clip]{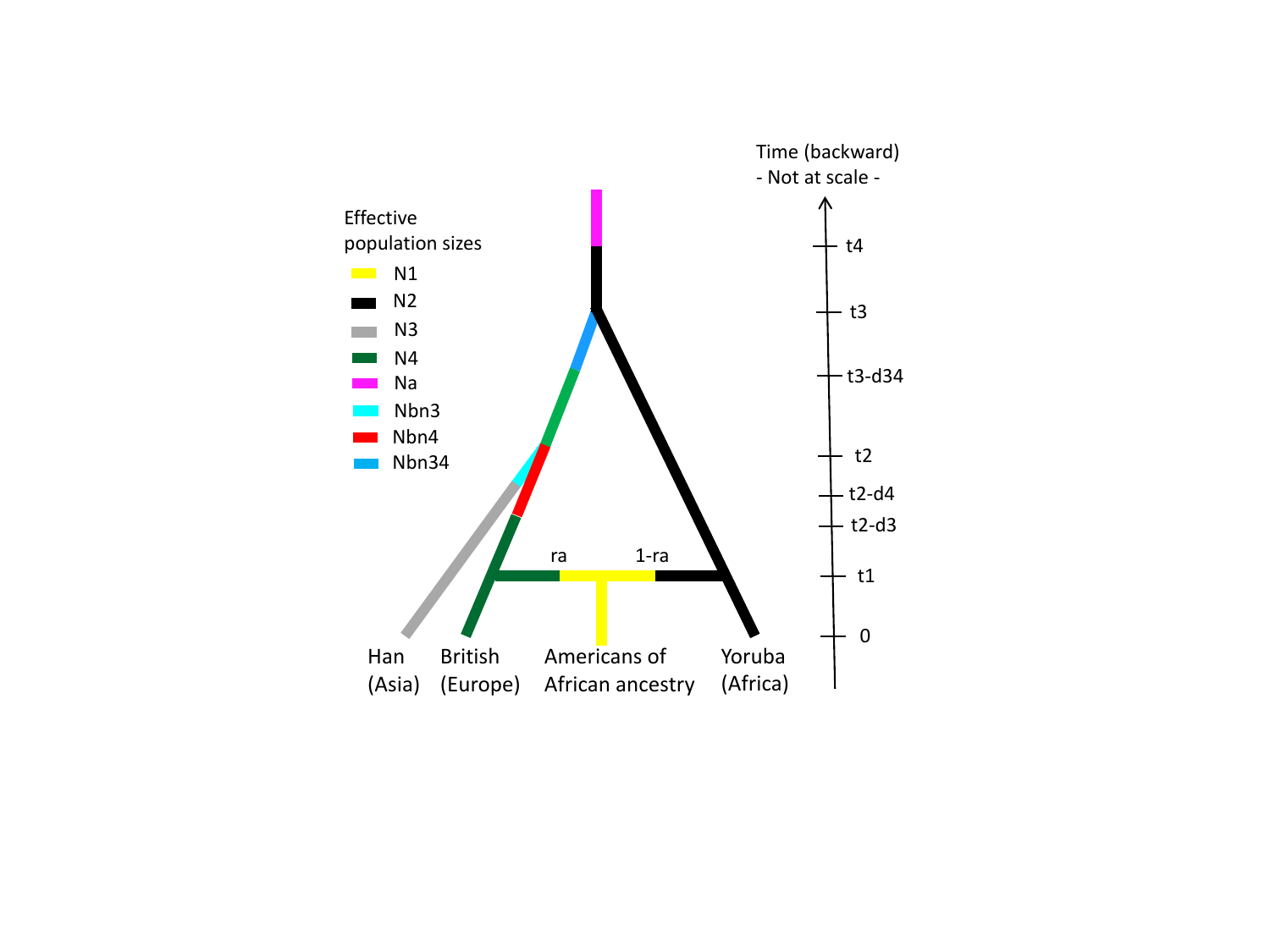}
  \caption{Evolutionary model of four human populations considered for Bayesian
parameter inference using ABC-RF. The prior distributions of the 
demographic and historical parameters
used to simulate SNP datasets are as followed: Uniform[100; 10\,000] for the
split times t2 and t3 (in number of generations), Uniform[1; 30] for the
admixture time t1, Uniform[0.05; 0.95] for the admixture rate ra (proportion of
genes with a non-African origin), Uniform[1000; 100\,000] for the stable
effective population sizes N1, N2, N3, N4 and N34 (in number of diploid
individuals), Uniform[5; 500] for the bottleneck effective population sizes
Nbn3, Nbn4, and Nbn34, Uniform[5; 500] for the bottleneck durations d3, d4, and
d34, Uniform[100; 10\,000] for both the ancestral effective population size Na
and t4 the time of change to Na. Conditions on time events were t4$>$t3$>$t2.
See \cite{pudlo:etal:2016} for details. Regarding the genetic model, we simulated biallelic
polymorphic SNP datasets using the algorithm proposed by \cite{hudson:2002} 
(cf ``-s 1'' option in the program \textit{ms} associated to \cite{hudson:2002}). 
This coalescent-based algorithm provides  the simulation efficiency and speed necessary
in the context of ABC, where large numbers of simulated 
datasets including numerous (statistically independent) SNP loci have to be generated 
(see Supplementary Appendix S1 of \cite{cornuet:etal:2014} for additional
comments on Hudson?s algorithm).
}
  \label{fig:human:scenario}
\end{figure}

We now illustrate our methodological findings with the study of a population
genetics dataset including $50\,000$ single nucleotide polymorphic (SNP)
markers genotyped in four human population samples
(\citeauthor{the1000genomes:2012}, \citeyear{the1000genomes:2012}; see details
in \citeauthor{pudlo:etal:2016}, \citeyear{pudlo:etal:2016}). The four
populations include Yoruba (Africa; YRI), Han (East Asia; CHB), British
(Europe; GBR) and American individuals of African ancestry (North America;
ASW).
The considered evolutionary model is represented in Figure
\ref{fig:human:scenario}. It includes a single out-of-Africa event with a
secondarily split into one European and one East Asian population lineage and a
recent genetic admixture of Afro-Americans with their African ancestors and with Europeans.
The model was robustly chosen as most appropriate among a set of eight evolutionary
models, when compared using ABC-RF for model choice in \cite{pudlo:etal:2016}.

We here focused our investigations on two parameters of interest in this model: (i)
the admixture rate ra (i.e. the proportion of genes with a non-African origin) that
describes the genetic admixture between individual of British and African
ancestry in Afro-Americans individuals; and (ii) the ratio N2/Na between
the ancestral effective population size Na and African N2 (in number of diploid
individuals), roughly describing the increase of African population size in the
past. Considering ratios of effective population sizes allows preventing identifiability issues of the model.

We used the software DIYABC v.2.0 \citep{cornuet:santos:beaumont:2008, cornuet:pudlo:veyssier:2014} to generate a \textit{reference table} of size $200\,000$, with $N=199\,000$ datasets being
used as training dataset and $N_{pred}=1000$ remaining as test datasets.
RFs are built in the same way as for our Normal example and make use of the
$k=112$ summary statistics provided for SNP markers by DIYABC, \citep[see][and 
the Section \ref{sec:SS} of the supplementary material]{pudlo:etal:2016}.

Due to the complexity of this model, the exact calculation of any posterior
quantity of interest is infeasible. To bypass this difficulty we compute NMAE
using simulated parameters from the test table, rather than targeted posterior
expectations. Here, $95\%$ credible intervals (CI) are deduced from
posterior quantile estimate of order $2.5\%$ and $97.5\%$.
Performances are measured via mean range and coverage, with coverage
corresponding to the percentage of rightly bounded parameters. For example a
$95\%$ CI should provide coverage equal to $95\%$ of the test table.

\begin{figure}
  \centering
  \includegraphics[width=0.55\linewidth]{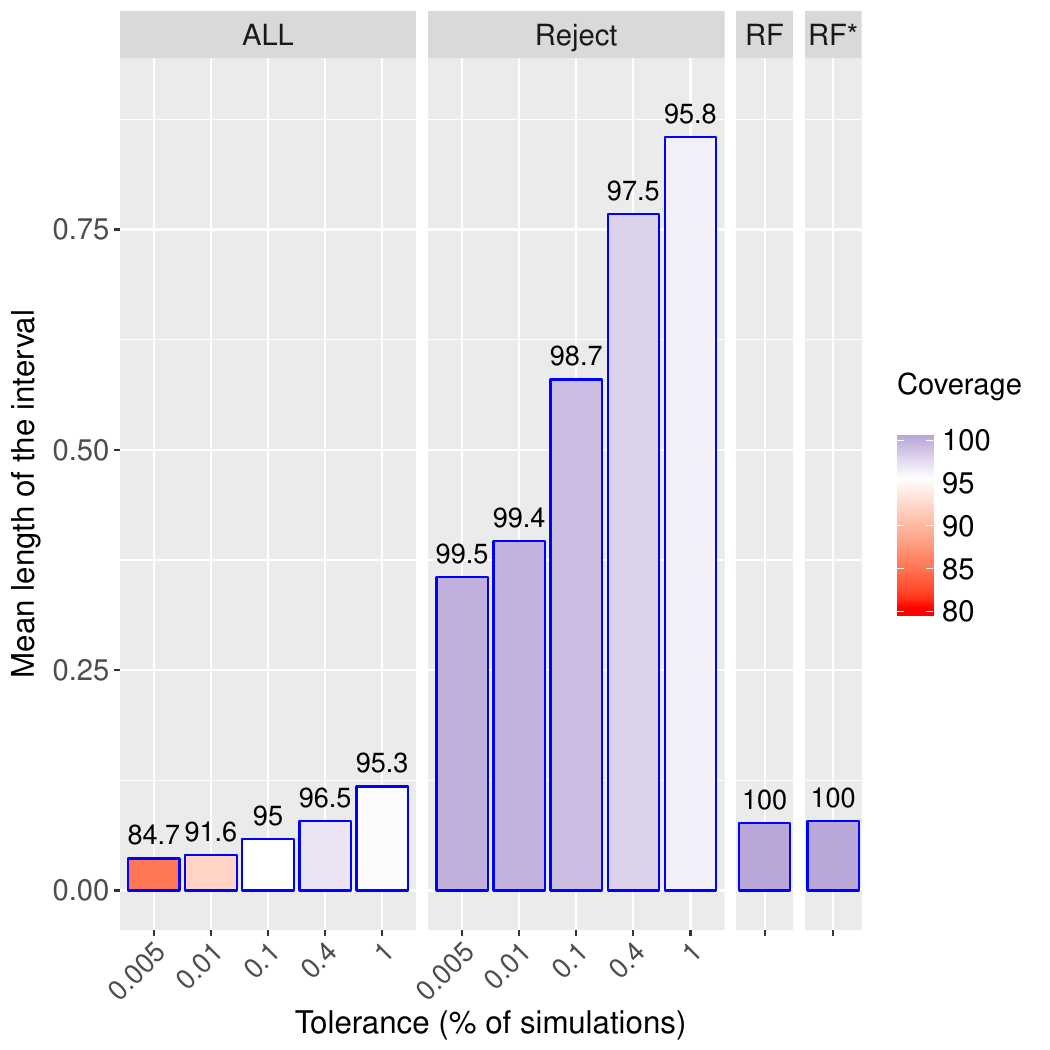}\\
  \includegraphics[width=0.55\linewidth]{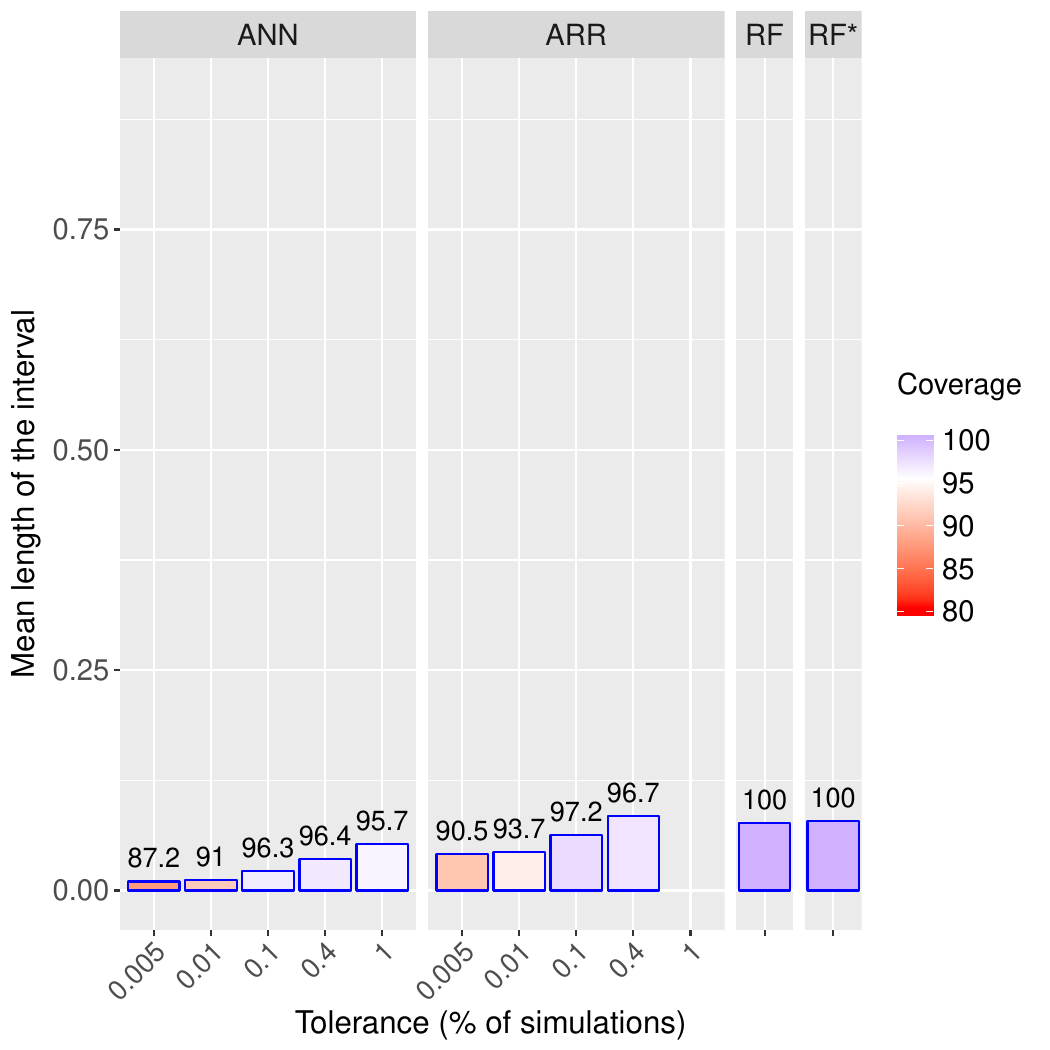}\\
    \caption{Range and coverage comparison of approximate $95\%$ credible
intervals on the admixture parameter ra (Figure \ref{fig:human:scenario})
obtained with ABC-RF (RF) and with earlier ABC methods : rejection (Reject), adjusted local linear (ALL) or ridge regression (ARR) or neural network (ANN) with various tolerance levels for Reject, ALL, ARR and ANN. Coverages values are specified by bar colors
and superimposed values. Heights indicate CI mean lengths. Results for ALL, Reject and RF are presented in the upper figure whereas those for ANN, ARR and RF are in the lower figure. See Figure \ref{fig:human:N2Na} for a similar representation of results for the parameter N2/Na. RF$^*$ refers to results obtained using ABC-RF when adding 20 additional independent noise variables generated from a uniform $\mathcal{U}_{[0,1]}$ distribution. RF refers to results without noise variables.}
  \label{fig:human:Ra}
\end{figure}

Figure \ref{fig:human:Ra} and Table \ref{tab:human:NMAE} (see also Figure \ref{fig:human:N2Na})
illustrate the quality of the ABC-RF method when compared with ABC with either
rejection, local linear, ridge or neural network adjustment (with logit
transforms of the parameters for non rejection methods) using different tolerance levels (i.e., with
tolerance proportion ranging from $0.005$ to $1$). We recall that considering
the ABC rejection method with a tolerance equals to 1 is equivalent to work
with the prior. Note that, due to memory allocation issues when using ABC method with adjusted ridge regression and a tolerance level of 1 on large \textit{reference table}, we did not manage to recover results in this specific case.

\begin{table}
\centerline{
\begin{tabular}{l c c c}
Method & Tolerance level & ra NMAE & N2/Na NMAE  \\
\hline
RF & NA & 0.018 & 0.053  \\
RF$^*$ & NA & 0.019 & 0.053  \\
Reject & 0.005 & 0.151 & 0.355 \\ 
Reject & 0.01 & 0.178 & 0.454 \\
Reject & 0.1 & 0.322 & 1.223 \\
Reject & 0.4 & 0.574 & 2.025 \\
Reject & 1 & 0.856 & 4.108 \\
ALL & 0.005 & 0.028 & 0.166 \\ 
ALL & 0.01 & 0.028 & 0.249 \\
ALL & 0.1 & 0.035 & 0.139 \\
ALL & 0.4 & 0.044 & 0.170 \\
ALL & 1 & 0.062 & 0.209 \\
ARR & 0.005 & 0.027 & 0.220 \\
ARR & 0.01 & 0.027 & 0.317 \\
ARR & 0.1 & 0.035 & 0.140 \\
ARR & 0.4 & 0.044 & 0.163 \\
ARR & 1 & $-$ & $-$ \\
ANN & 0.005 & \textbf{0.007} & \textbf{0.037} \\
ANN & 0.01 & 0.007 & 0.038 \\
ANN & 0.1 & 0.013 & 0.064 \\
ANN & 0.4 & 0.016 & 0.123 \\
ANN & 1 & 0.025 & 0.095
\end{tabular}
}
\caption{\label{tab:human:NMAE} Comparison of normalized mean absolute errors
(NMAE) for the estimation of the parameters ra and N2/Na using ABC-RF (RF) and
ABC with rejection (Reject), adjusted local linear (ALL) or ridge regression (ARR) or neural network (ANN) with various tolerance levels for Reject, ALL, ARR and ANN. NA stands for not appropriate. The smallest NMAE values are in by bold characters. NA stands for not appropriate. RF$^*$ refers to results obtained using ABC-RF when adding 20 additional independent noise variables generated from a uniform $\mathcal{U}_{[0,1]}$ distribution. RF refers to results without noise variables.}
\end{table}
Interesting methodological features can be observed in association with this example. ABC with rejection performs poorly in terms of NMAE and provides conservative and hence 
wide CIs (i.e., with coverage higher than the formal level). For ABC with 
adjustment, the lower the tolerance the lower the error (Table
\ref{tab:human:NMAE}). The CI quality however highly suffers from low
tolerance, with underestimated coverage (Figures \ref{fig:human:Ra} and
\ref{fig:human:N2Na}). The smaller the tolerance value, the narrower the CI. Results for the ABC method with adjusted ridge regression seems however to be unstable for the parameter N2/Na depending on the considered level of tolerance.
The ABC method using neural network and a tolerance level of $0.005$ provides the lowest NMAE for both parameters of interest. The corresponding coverages are however underestimated, equal to $87.2\%$ for ra and $81.6\%$ for N2/Na, when $95\%$ is expected (lower part of Figure \ref{fig:human:Ra} and Figure \ref{fig:human:N2Na}). Note that results with this method can be very time consuming to obtain when the tolerance level and the number of layers are large.
The ABC-RF method provides an appealing trade-off between parameter estimation quality (ABC-RF is the method with the second lowest NMAE values in Table \ref{tab:human:NMAE}) and
slightly conservative CIs (Figures \ref{fig:human:Ra} and \ref{fig:human:N2Na}).
Similar results and methodological features were observed when focusing on the $90\%$ CI (results not shown). It is also worth
stressing that not any calibration of any kind of a tolerance level parameter are needed with
ABC-RF, which is an important plus for this method. On the opposite, earlier ABC methods require calibration to optimize their use, such calibration being time consuming when different levels of tolerance are used.

For the observed dataset used in this study, posterior expectations and quantiles
of the parameters of interest ra and N2/Na are reported in Table
\ref{tab:human:obs}. Expectation and CI values substantially vary for both
parameters, depending on the method used. The impact of the tolerance levels is
noteworthy for both the rejection and local linear adjustment ABC methods. The
posterior expectation of ra obtained using ABC-RF was equal to $0.221$ with a
relatively narrow associated $95\%$ CI of $[0.112, 0.287]$. The latter
estimation lays well within previous estimates of the mean proportion of genes
of European ancestry within African American individuals, which typically
ranged from $0.070$ to $0.270$ $-$ with most estimates around $0.200$ $-$,
depending on individual exclusions, the population samples and sets of genetic
markers considered, as well as the evolutionary models assumed and inferential
methods used \citep[reviewed in][]{bryc:etal:2015}. 
Interestingly, a recent
genomic analysis using a conditional random field parametrized by random
forests trained on reference panels \citep{maples:etal:2013} and $500\,000$
SNPs provided a similar expectation value of ra for the same African American
population ASW (i.e. ra $= 0.213$), with a somewhat smaller $95\%$ CI (i.e.
$[0.195, 0.232]$), probably due to the ten times larger number of SNPs in their dataset \citep{baharian:etal:2016}.

The posterior expectation of N2/Na obtained using ABC-RF was equal to $4.508$
with a narrow associated $95\%$ CI of $[3.831, 5.424]$. Such values point to
the occurrence of the substantial ancestral demographic and geographic expansion that
is widely illustrated in previous Human population genetics studies, including African populations \citep[e.g.][]{henn:etal:2012}. Although our modeling setting assumes a
na{\"\i}ve abrupt change in effective population sizes in the ancestral African
population, the equivalent of N2/Na values inferred from different methods and
modeling settings fit rather well with our own posterior expectations and
quantiles for this parameter \citep[e.g.][]{schiffels:durbin:2014}.

\begin{table}
\footnotesize
\centerline{
\begin{tabular}{l c c c c c c}
\multicolumn{7}{c}{ra} \\
\midrule
Method & Tol. level & Expectation & $Q_{0.025}$ & $Q_{0.05}$ & $Q_{0.95}$ & $Q_{0.975}$   \\
\hline
RF & NA & 0.221 & 0.112 & 0.134 & 0.279 & 0.287  \\
RF$^*$ & NA & 0.225 & 0.112 & 0.142 & 0.282 & 0.290 \\
Reject & 0.005 & 0.223 & 0.061 & 0.069 & 0.364 & 0.389 \\ 
Reject & 0.01 & 0.220 & 0.060 & 0.070 & 0.389 & 0.418 \\
Reject & 0.1 & 0.276 & 0.062 & 0.074 & 0.511 & 0.543 \\
Reject & 0.4 & 0.388 & 0.068 & 0.086 & 0.739 & 0.791 \\
Reject & 1 & 0.502 & 0.073 & 0.095 & 0.906 & 0.928  \\
ALL & 0.005 & 0.278 & 0.219 & 0.229 & 0.322 & 0.337 \\ 
ALL & 0.01 & 0.257 & 0.232 & 0.238 & 0.274 & 0.278 \\
ALL & 0.1 & 0.207 & 0.170 & 0.171 & 0.233 & 0.237 \\
ALL & 0.4 & 0.194 & 0.144 & 0.152 & 0.233 & 0.241 \\
ALL & 1 & 0.196 & 0.115 & 0.126 & 0.278 & 0.299 \\
ARR & 0.005 & 0.260 & 0.252 & 0.254 & 0.265 & 0.266 \\ 
ARR & 0.01 & 0.252 & 0.239 & 0.242 & 0.260 & 0.262 \\
ARR & 0.1 & 0.211 & 0.171 & 0.178 & 0.239 & 0.244 \\
ARR & 0.4 & 0.196 & 0.140 & 0.149 & 0.241 & 0.251 \\
ARR & 1 & $-$ & $-$ & $-$ & $-$ & $-$ \\
ANN & 0.005 & 0.227 & 0.221 & 0.223 & 0.232 & 0.234 \\
ANN & 0.01 & 0.226 & 0.219 & 0.221 & 0.231 & 0.233 \\
ANN & 0.1 & 0.228 & 0.217 & 0.220 & 0.236 & 0.239\\
ANN & 0.4 & 0.232 & 0.216 & 0.221 & 0.242 & 0.248 \\
ANN & 1 & 0.206 & 0.183 & 0.187 & 0.227 & 0.233 \\
\multicolumn{7}{c}{} \\
\multicolumn{7}{c}{N2/Na} \\
\midrule
Method & Tol. level & Expectation & $Q_{0.025}$ & $Q_{0.05}$ & $Q_{0.95}$ & $Q_{0.975}$   \\
\hline
RF & NA & 4.508 & 3.831 & 3.959 & 5.153 & 5.424 \\
RF$^*$ & NA & 4.594 & 3.821 & 3.910 & 5.241 & 6.552 \\
Reject & 0.005 & 6.282 & 2.937 & 3.223 & 10.086 & 11.337 \\ 
Reject & 0.01 & 6.542 & 2.746 & 3.116 & 10.837 & 11.852 \\
Reject & 0.1 & 8.001 & 2.131 & 2.574 & 15.690 & 18.531 \\
Reject & 0.4 & 11.605 & 1.795 & 2.331 & 28.011 & 38.532 \\
Reject & 1 & 23.483 & 0.672 & 1.185 & 84.649 & 147.657 \\
ALL & 0.005 & 30.041 & 1.256 & 1.879 & 83.369 & 174.340 \\ 
ALL & 0.01 & 9.289 & 3.946 & 4.586 & 16.686 & 20.361 \\
ALL & 0.1 & 8.235 & 5.736 & 5.995 & 11.573 & 12.719 \\
ALL & 0.4 & 10.752 & 4.588 & 4.996 & 21.656 & 27.300 \\
ALL & 1 & 7.222 & 5.684 & 5.829 & 9.631 & 10.475 \\
ARR & 0.005 & 10.528 & 4.395 & 5.677 & 19.224 & 22.722 \\
ARR & 0.01 & 8.264 & 5.020 & 5.485 & 12.544 & 13.313 \\
ARR & 0.1 & 8.394 & 5.643 & 5.948 & 12.075 & 13.313 \\
ARR & 0.4 & 10.802 & 6.113 & 6.505 & 17.487 & 20.511 \\
ARR & 1 & $-$ & $-$ & $-$ & $-$ & $-$  \\
ANN & 0.005 & 5.746 & 5.512 & 5.563 & 5.937 &  5.982 \\
ANN & 0.01 & 6.148 & 5.883 & 5.934 & 6.353 & 6.420 \\
ANN & 0.1 & 25.921 & 23.857 & 24.250 & 27.672 & 28.133 \\
ANN & 0.4 & 8.515 & 7.652 & 7.810 & 9.147 & 9.436 \\
ANN & 1 & 7.021 & 5.692 & 5.856 & 8.677 & 9.370 
\end{tabular}
}
\caption{\label{tab:human:obs} Estimation of the two parameters of interest ra
and N2/Na for the observed human population genetics dataset using ABC-RF (RF), and ABC with rejection
(Reject), adjusted local linear (ALL) or ridge regression (ARR) or neural network (ANN) with various tolerance levels (Tol. level) for Reject, ALL, ARR and ANN. NA stands for not appropriate. RF$^*$ refers to results obtained using ABC-RF when adding 20 additional independent noise variables generated from a uniform $\mathcal{U}_{[0,1]}$ distribution. RF refers to results without noise variables.}
\end{table}

In contrast to earlier ABC methods, the RF approach is deemed to be mostly insensitive to the presence of covariates whose the distributions does not depend on the parameter values (i.e. ancillary covariates) \citep[e.g.][]{breiman:2001, marin:pudlo:robert:ryder:2012}. To illustrate this feature, we have added $20$ additional independent noise variables generated from a uniform $\mathcal{U}_{[0,1]}$ distribution (results designated by RF$^*$) in the \textit{reference table} generated for the present Human population genetics example. We found that the presence of such noise covariates do not impact the results in terms of NMAE, coverage and only slightly on parameter estimation for the observed dataset (Tables \ref{tab:human:NMAE} and \ref{tab:human:obs}, and Figures \ref{fig:human:Ra} and \ref{fig:human:N2Na}). For the rest of the article, no noise variables were used.

\subsection{Practical recommendations regarding the implementation of the ABC-RF algorithm}
\label{subsec:practical}

We mainly consider in this section two important practical issues, namely the choice
of the number of simulations ($N$) in the \textit{reference table} and of the
number of trees ($B$) in the random forest. For sake of simplicity and concision, we focus our recommendations on the
above human population genetics example (subsection \ref{subsec:human}). We stress here that, although not generic, our recommendations fit well with other examples of complex model settings that we have analysed so far (results not shown). We also stress that for simpler model settings substantially smaller $N$ and $B$ values were sufficient to obtain good results. Finally, we provide practical comments about the main sources of variabilities in inferences typical of the ABC-RF methodology.

\paragraph{\textbf{\textit{Reference table} size $-$}}
We consider a reference table made of $N=199\,000$ simulated datasets.
However, Table \ref{tab:recommend:NMAE} shows a negligible decrease of NMAE
when using $N=100\,000$ to $N=199\,000$ datasets. Table \ref{tab:recommend:pred} also exhibits small variations 
between predictions on the observed dataset, especially for $N \geq 7500$. The level of variation thus seems
to be compatible with the random variability of the RF themselves. Altogether, using a
\textit{reference table} including $100\,000$ datasets seems to be a reasonable default
choice. It is worth stressing that the out-of-bag mean squared error can be easily retrieved and provides a good
indicator of the quality of the RF without requiring the simulation of a (small size) secondary test table, which can hence be estimated at a low computational cost (Table \ref{tab:recommend:NMAE}).

\begin{table}
\centerline{
\begin{tabular}{l c c c c c c c}
\multicolumn{8}{c}{NMAE} \\
\midrule
$N$ $(\times 10^3)$ & $10$ & $25$ & $50$ & $75$ & $100$ & $150$ & $199$ \\
\hline
ra & 0.028 & 0.023 & 0.021 & 0.020 & 0.019 & 0.018 & 0.018 \\
N2/Na & 0.080 & 0.067 & 0.059 & 0.057 & 0.055 & 0.053 & 0.053 \\
\multicolumn{8}{c}{} \\
\multicolumn{8}{c}{OOB MSE} \\
\midrule
$N$ $(\times 10^3)$ & $10$ & $25$ & $50$ & $75$ & $100$ & $150$ & $199$ \\
\hline
ra $(\times 10^{-4})$ & 1.670 & 1.176 & 0.914 & 0.823 & 0.745 & 0.695 & 0.664 \\
N2/Na $(\times 10^{3})$ & 0.194 & 0.179 & 0.143 & 0.125 & 0.115 & 0.111 & 0.110 \\
\end{tabular}
}
\caption{\label{tab:recommend:NMAE} Comparison of normalized mean absolute errors (NMAE) and out-of-bag mean squared errors (OOB MSE) for the estimation of the parameters ra and N2/Na obtained with ABC-RF, using different \textit{reference table} sizes ($N$). We use the test table mentioned in subsection \ref{subsec:human}. The number of trees in the RF is 500.}
\end{table}

\begin{table}
\centerline{
\begin{tabular}{l c c c c c c c}
$N$ $(\times 10^3)$ & $10$ & $25$ & $50$ & $75$ & $100$ & $150$ & $199$ \\
\hline
ra expectation & 0.231 & 0.222 & 0.224 & 0.223 & 0.222 & 0.223 & 0.221 \\
ra $Q_{0.025}$ & 0.097 & 0.095 & 0.102 & 0.104 & 0.106 & 0.109 & 0.112 \\
ra $Q_{0.975}$ & 0.317 & 0.309 & 0.305 & 0.305 & 0.289 & 0.292 & 0.287 \\
\hline
N2/Na expectation & 4.538 & 4.588 & 4.652 & 4.530 & 4.475 & 4.483 & 4.508 \\
N2/Na $Q_{0.025}$ & 3.651 & 3.679 & 3.782 & 3.802 & 3.751 & 3.840 & 3.831 \\
N2/Na $Q_{0.975}$ & 6.621 & 6.221 & 6.621 & 5.611 & 5.555 & 5.315 & 5.424 \\
\end{tabular}
}
\caption{\label{tab:recommend:pred} Estimation of the parameters ra and N2/Na for the observed population genetics dataset with ABC-RF, using different \textit{reference table} sizes ($N$). The number of trees in the RF is 500.}
\end{table}

\begin{figure}
  \centering
  \includegraphics[width=0.7\linewidth]{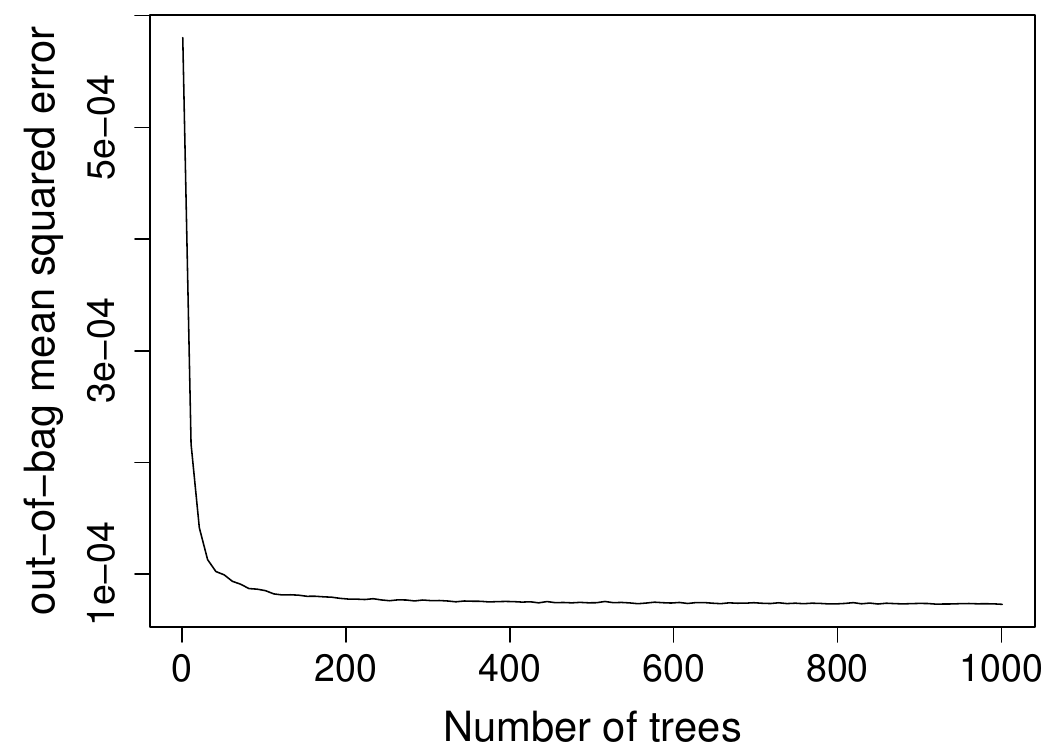}\\
  \caption{Relations between the number of trees in the forest and the ABC-RF out-of-bag mean squared errors, for a reference table of size $N=100\,000$ in the human population genetics example.}
  \label{fig:human:nbrTrees}
\end{figure}

\paragraph{\textbf{Number of trees $-$}} A forest including 500 trees is a
default choice when building RFs, as this provides a good trade-off between
computation efficiency and statistical precision \citep{breiman:2001,pudlo:etal:2016}.
To evaluate whether or not this number is sufficient, we recommend to compute
the out-of-bag mean squared error depending on the number of trees in the
forest for a given \textit{reference table}. If 500 trees is a satisfactory
calibration, one should observe a stabilization of the error around this value.
Figure \ref{fig:human:nbrTrees} illustrates this representation on the human
population genetics example and points to a negligible decrease
of the error after 500 trees. This graphical representation is produced via
our {\sf R} package \texttt{abcrf}.
\
\paragraph{\textbf{Minimum node size (maximum leaf size) $-$}} We recall that splitting events during a tree construction stop when a node has less than $N_\text{min}$ observations,
in that case, the node becomes a leaf. Note that the higher  $N_\text{min}$  the quicker RF treatments. In all RF treatments presented here, we used the default size $N_\text{min}=5$. Table \ref{tab:recommend:node:size} illustrates the influence of $N_\text{min}$ on the human population genetics example and highlights a negligible decrease of the error for $N_\text{min}$ lower than $5$.

\begin{table}
\centerline{
\begin{tabular}{l c c c c c c c c c c c}
\multicolumn{12}{c}{NMAE} \\
\midrule
$N_\text{min}$ & $1$ & $2$ & $3$ & $4$ & $5$ & $10$ & $20$ & $50$& $100$& $200$& $500$ \\
\hline
ra & 0.019 & 0.019 & 0.019 & 0.019 & 0.019 & 0.019 & 0.020 & 0.021 & 0.023 & 0.027 & 0.033  \\
N2/Na & 0.054 & 0.055 &0.054 &0.055 &0.055 &0.055 & 0.055 & 0.058 & 0.062 & 0.068 & 0.082\\
\multicolumn{12}{c}{} \\
\multicolumn{12}{c}{OOB MSE} \\
\midrule
$N_\text{min}$ & $1$ & $2$ & $3$ & $4$ & $5$ & $10$ & $20$ & $50$& $100$ & $200$ & $500$ \\
\hline
ra $(\times 10^{-4})$ & 0.745 & 0.739 & 0.744 & 0.739 & 0.745 & 0.760 & 0.783 & 0.925 & 1.129 & 1.480 & 2.280 \\
N2/Na $(\times 10^{3})$ & 0.114 & 0.116 & 0.115 & 0.115 & 0.115 & 0.116 & 0.119 & 0.131 & 0.153 & 0.183 & 0.252 \\
\end{tabular}
}
\caption{\label{tab:recommend:node:size} Comparison of normalized mean absolute errors (NMAE) and out-of-bag mean squared errors (OOB MSE) for the estimation of the parameters ra and N2/Na obtained with ABC-RF, using different minimum node sizes ($N_\text{min}$). We use the reference table of size $N=100\,000$ and the test table mentioned in subsection \ref{subsec:human}. The number of trees in the RF is 500.}
\end{table}

Finally, we see no reason to change the number of summary statistics sampled at each split $n_\text{try}$ within a tree, which is traditionally chosen as $k/3$ for regression when $k$ is the total number of predictors \citep{breiman:2001}.

\paragraph{\textbf{Variability in the ABC-RF methodology $-$}}
The ABC-RF methodology is associated with different sources of variabilities the user should be aware of. Using a simulated \textit{reference table} is the main source, RF being the second. Indeed, predicting quantities of interest for the same test dataset with two different \textit{reference tables} of equal size $N$ will result in slightly different estimates. This variation has been previously highlighted in Figure \ref{fig:true_vs_RFcdfs} dealing with the analysis of the Normal toy example. We recall RF are composed of trees trained on bootstrap samples, each one considering $n_\text{try}$ covariates randomly selected amongst the $k$ available at each split. This random aspects of RF results in variability.
In practice, a good user habit should be to run ABC-RF more than once on different training datasets to ensure that the previously mentioned variabilities are negligible. If this variability is significant, we recommend considering a \textit{reference table} of higher size.

\section{Discussion}

This paper introduces a novel approach to parameter estimation in
likelihood-free problems, relying on the machine-learning tool of regression RF
to automatise the inclusion of summary statistics in ABC algorithms. Our
simulation experiments demonstrate several advantages of our methodological proposal compared with earlier ABC methods.

While using the same \textit{reference table} and test dataset for all compared methods, our RF
approach appears to be more accurate than previous ABC solutions.
Approximations of expectations are quite accurate, while posterior variances are only
slightly overestimated, which is an improvement compared with other approaches
that typically underestimate these posterior variances. The performances for covariance approximation are
quite encouraging as well, although the method is still incomplete and need
further developments on this particular point (more details are given in Section \ref{sup:sec:cov} of the supplementary material). We found that quantile
estimations depend on the corresponding probability and we believe this must be related to the approximation error of
the posterior cumulative function $F(x \mid \eta(y))$. More specifically, we
observed that upper quantiles may be overestimated, whereas lower quantiles
may be underestimated (Figure \ref{fig:QQplot-quantiles}), indicating fatter tails in the approximation. Hence,
credible intervals produced by the RF solution may be larger than the exact
ones. However from a risk assessment point of view, this overestimation aspect
clearly presents less drawbacks than underestimation of credible intervals. Altogether, owing to the various models and datasets we analysed,
we argue that ABC-RF provides a good trade-off in terms of quality
between parameter estimation of point estimators (e.g. expectation, median or variance) and credible interval coverage. 
A comparison of computing times is given in Section \ref{sup:temps} of the supplementary material.

Throughout our experiments, we found that, contrary to earlier ABC methods, the RF approach is mostly
insensitive to the presence of covariates whose the distributions does not depend on the parameter values (ancillary covariates). 
Therefore, we argue that the RF method can deal with a very large number of summary statistics, bypassing any form of
pre-selection of those summaries. 
Interestingly, the property of
ABC-RF to extract and adaptively weight information carried by each of the
numerous summary statistics proposed as explanatory variables can be
represented by graphs, showing the relative contribution of summary statistics
in ABC-RF estimation for each studied parameter (see Section \ref{sup:importance} of the supplementary material for details).

As an alternative, \cite{papamakarios:murray:2016} propose to approximate the whole posterior distribution by using Mixture Density Networks \citep[MDN,][]{bishop:1994}. The MDN strategy consists in using Gaussian mixture models with parameters calibrated thanks to neural networks. The strategy of \cite{papamakarios:murray:2016} is to iteratively learn an efficient proposal prior (approximating the posterior distribution), then to use this proposal to train the posterior, both steps making use of MDN. This strategy can be easily applied when the prior is uniform or Gaussian, but other prior choices
can involve difficulties. This is because in such cases, it might be difficult to simulate from the corresponding proposal.
The approximation accuracy of the posterior as a Gaussian mixture model depends of the number of components and the number of hidden layers of the networks. Those two parameters require calibration. Finally, by using MDN, one loses
the contribution of summary statistics provided by RF and thus some useful interpretation
elements. Despite these remarks, this promising method remains of interest and is worth mentioning.

The RF method focuses on unidimensional parameter inference. Multi-objective random forest \citep{kocev:etal:2007} could be a solution to deal with multidimensional parameter using RF. However, our attempts based on the later methodology were so far unfruitful (results not shown). An alternative approach could be based on using the RF strategies to approximate some conditionals distributions
and then recover the joined posterior using either a Gibbs sampler (based on approximated full conditionals) or a Russian rule
decompositions to which a product of embedded full conditionals is associated. We are presently comparing the two strategies 
on simulated datasets.

In population genetics, which historically corresponds to the field of 
introduction of ABC methods, next generation sequencing technologies result in
large genome-wide datasets that can be quite informative about the demographic
history of the genotyped populations. Several recently developed inferential
methods relying on the observed site frequency spectrum appear particularly
well suited to accurately characterizing the complex evolutionary history of invasive
populations \citep{gutenkunst:etal:2009, excoffier:etal:2013}.
Because of the reduced computational resources demanded by ABC-RF and the
above-mentioned properties of the method, we believe that ABC-RF can efficiently contribute
to the analysis of massive SNP datasets, including both model choice
\citep{pudlo:etal:2016} and Bayesian inference about parameters of interest. More
generally, the method should appeal to all scientific fields in which large
datasets and complex models are analysed using simulation-based methods such as
ABC \citep[e.g.][]{beaumont:2010, sisson:etal:2017}.

\section*{Acknowledgements}

The use of random forests was suggested to J.-M.M. and C.P.R. by Bin Yu during
a visit at CREST, Paris. We are very grateful to her for this conversation and also to Nicolai
Meinshausen for some very helpful exchanges on the quantile regression forest
method. We thank Dennis Prangle and Michael Blum for their very helpful suggestions, helping us to improve the quality of that paper.

\section*{Funding}
A.E. acknowledges financial support by the National Research Fund ANR (France) through the European Union program ERA-Net BiodivERsA (project EXOTIC) and the project ANR-16-CE02-0015-01 (SWING), and the INRA scientific department SPE (AAP-SPE 2016). C.P.R. is supported by an IUF 2016--2021 senior grant.

\newpage

\begin{center}
\textbf{\huge Supplementary Information \\ ABC random forests for Bayesian parameter inference}
\end{center}
\setcounter{equation}{0}
\setcounter{figure}{0}
\setcounter{table}{0}
\setcounter{page}{1}
\setcounter{section}{0}

\makeatletter
\renewcommand{\theequation}{S\arabic{equation}}
\renewcommand{\thefigure}{S\arabic{figure}}
\renewcommand{\thetable}{S\arabic{table}}
\renewcommand{\bibnumfmt}[1]{[S#1]}
\renewcommand{\citenumfont}[1]{S#1}

\vspace{3cm} \section{Comparing three methods of variance estimation of parameters}
\label{threeMethods}

We here compare three methods to estimate posterior variance of a parameter transformation of interest $\tau$ using ABC-RF. Two of them have already been explained in the main text (i.e. method 1 and 3 below).
\begin{itemize}

\item \textbf{Method 1}: One reuses the original random forest (RF) weights $w_t(\eta(y))$ to the out-of-bag square residuals $(\tau^{(t)} - \hat{\tau}_{ \text{oob} }^{(t)})^2$, giving the variance estimator
\begin{align*}
\widetilde{ \mbox{Var} } (\tau \mid \eta(y)) = \sum_{t=1}^N w_t(\eta(y)) (\tau^{(t)} - \hat{\tau}_{ \text{oob} }^{(t)})^2.
\end{align*}

\item \textbf{Method 2}: A similar estimator can be obtained by building a new RF thanks to the training sample  $(\eta(y^{(t)}), \big(\tau^{(t)} - \tau_\text{oob})^2 \big)_{t=1,\ldots,N}$,
resulting in the estimator
\begin{align*}
\mbox{Var}^\#(\tau \mid \eta(y) ) = \sum_{t=1}^N \widetilde{w_t}(\eta(y)) (\tau^{(t)} - \hat{\tau}_{ \text{oob} }^{(t)})^2,
\end{align*}
where $\widetilde{w_t}(\eta(y))$ is the computed weights of this newly trained RF. This estimator is based on the expression of the  posterior variance as a conditional expectation: $$\mbox{Var}(\tau \mid \eta(y)) = \mathbb{E}\left( \left[ \tau - \mathbb{E}(\tau \mid \eta(y)) \right]^2 \mid \eta(y) \right)$$ and the fact that such a RF is able to estimate this posterior expectation. 
This approach is more expensive due to the additional RF requirement.

\item \textbf{Method 3}: The variance estimator is based on the cumulative distribution function (cdf) approximation,
\begin{align*}
\widehat{\mbox{Var}}(\tau \mid \eta(y) ) = \sum_{t=1}^N w_t(\eta(y)) \left( \tau^{(t)} -  \sum_{u=1}^N w_u(\eta(y)) \tau^{(u)} \right)^2.
\end{align*}

\end{itemize}

\begin{figure}[H]
  \centering
  \includegraphics[width=0.49\linewidth]{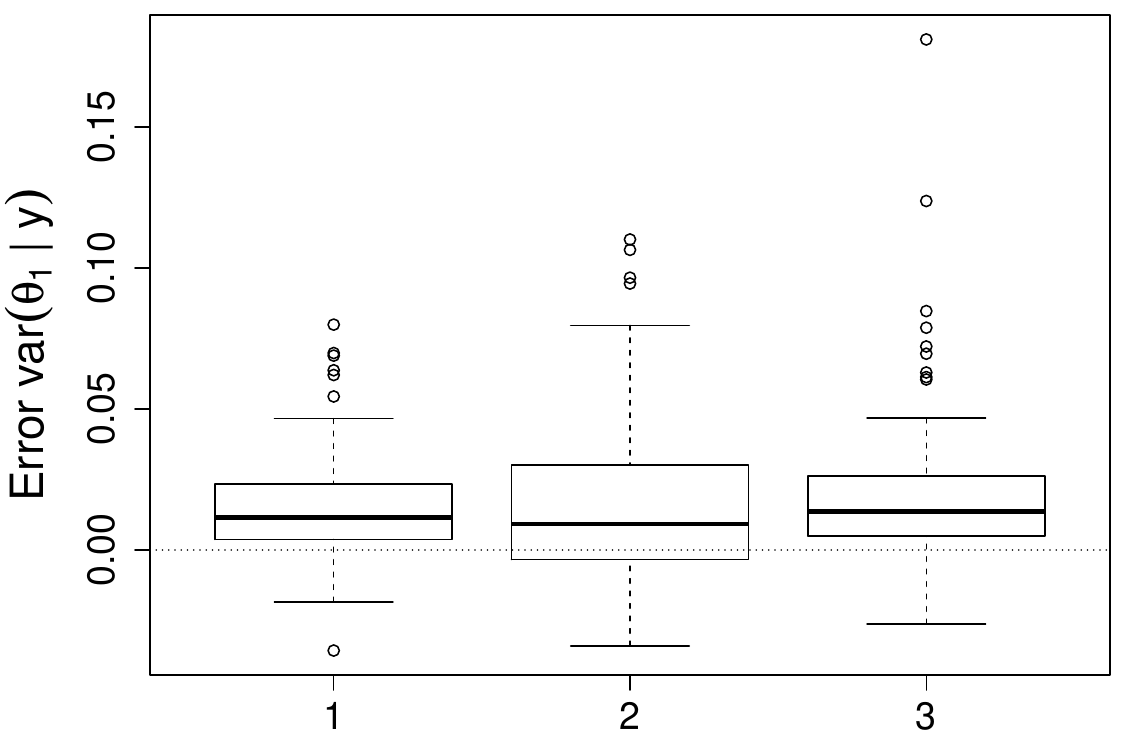}
  \includegraphics[width=0.49\linewidth]{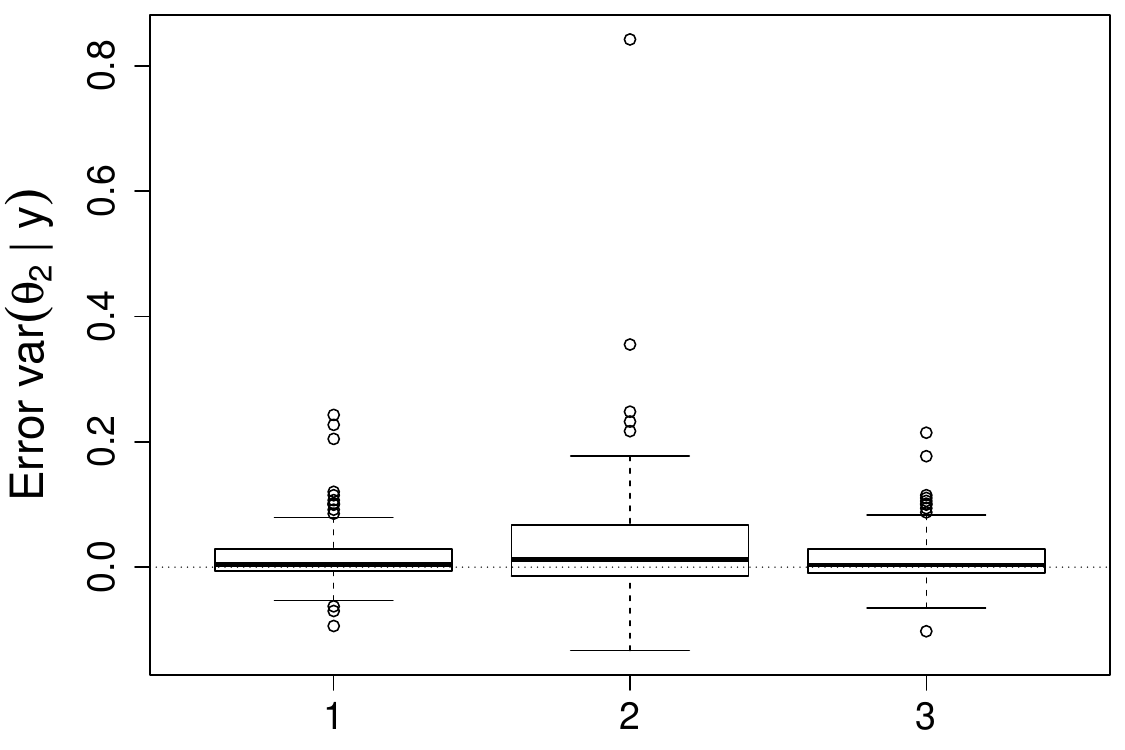}
  \caption{Boxplot comparison of differences between our predictions for $\mbox{Var}(\theta_i \mid y)$ and the true values with the three methods of variance estimation: by reusing weights (method 1, boxplot 1), by building a new RF on square residuals (method 2, boxplot 2) and by using the estimation of the cumulative distribution function (method 3, boxplot 3).}
  \label{fig:boxplot-variances}
\end{figure}

We here compare these three estimators on the Normal toy example detailed in the main text with $h$ the projection on both coordinates of the parameter vector $\theta$ (see Section 3.1 of the main text).
We find that the three estimators behave similarly; boxplots are alike and all tend to overestimate posterior variances (Figure \ref{fig:boxplot-variances}). Results summarized in Table \ref{tab:gaus-NMAE-three-methods} also supports this similarity in NMAE terms. Because the estimator 1 appears to show slightly lower errors for both parameter $\theta_1$ and $\theta_2$, we decided to use it in the two examples detailed in the main text.

\vspace{2cm}

\begin{table}[H]
\centerline{
\begin{tabular}{c c c c}
Method & 1 & 2 & 3  \\
\hline
$\hbox{Var}(\theta_1 \mid y)$ & \textbf{0.25} & 0.28 & 0.30  \\
$\hbox{Var}(\theta_2 \mid y)$ & \textbf{0.25} & 0.47 & \textbf{0.25} \\
\end{tabular}
}
\caption{\label{tab:gaus-NMAE-three-methods} Comparison of normalized mean absolute errors (NMAE) of estimate variances when using three methods, (see legend of Figure \ref{fig:boxplot-variances} and Section \ref{threeMethods} of the supplementary material). The smallest NMAE values are in bold characters.}
\end{table}

\clearpage

\section{Study of covariances of parameters using random forests}
\label{sup:sec:cov}

\subsection{Methodology}

We are here interested in another estimate that is frequently produced in a Bayesian analysis, that is the posterior covariance between two univariate transforms of the parameter, $\tau=h(\theta)$ and $\sigma=g(\theta)$ say, 
$\mbox{Cov}(\tau, \sigma \mid \eta(y))$. Since
we cannot derive this quantity from the approximations to the marginal posteriors of $\tau$ and $\sigma$, we propose to construct a specific RF for this purpose. With the same notations used in the main text, we denote approximations of posterior expectations for $\tau$ and $\sigma$, produced by out-of-bag informations, by $\hat{\tau}_{ \text{oob} }^{(t)}$ and $\hat{\sigma}_{ \text{oob} }^{(t)}$. We use the product of out-of-bag errors for $\tau$ and $\sigma$ in the empirical covariance, and consider
$(\tau^{(t)}- \hat{\tau}_{ \text{oob} }^{(t)} ) (\sigma^{(t)}- \hat{\sigma}_{ \text{oob} }^{(t)} )$ as the response variable. With the previously introduced
notations, the corresponding RF estimator is
\begin{equation*}
\widetilde{ \mbox{Cov} } (\tau, \sigma \mid \eta(y)) = \frac{1}{B} \sum_{b=1}^B \frac{1}{\big|L_b(\eta(y))\big|} \sum_{t : \eta(y^{(t)}) \in L_b} n_b^{(t)}
(\tau^{(t)}- \hat{\tau}_{ \text{oob} }^{(t)} ) (\sigma^{(t)}- \hat{\sigma}_{ \text{oob} }^{(t)} ).
\end{equation*}
This posterior covariance approximation requires a total of three regression RFs: one for each parameters and one for the covariance approximation.

\subsection{Toy regression example}

We now apply our RF methodology to a toy regression example for which its non-zero covariance between parameters is the main quantity of interest, hence we consider the case were $g$ and $h$ are the projections on a given coordinate of the parameter vector $\theta$. For a simulated $n \times 2$ design matrix $X=[x_1,x_2]$, we consider the Zellner's hierarchical model \citep[chapter 3]{sup:marin:robert:2014}
\begin{align*}
(y_1, \ldots, y_n) \mid \beta_1,\beta_2, \sigma^2 &\sim \mathcal{N}_n(X\beta, \sigma^2Id), \\
\beta_1, \beta_2 \mid \sigma^2 &\sim \mathcal{N}_2(0, n\sigma^2(X^\top X)^{-1}), \\
\sigma^2 &\sim I\mathcal{G}(4,3),
\end{align*}
where $\mathcal{N}_k \left(\mu, \Sigma \right)$ denotes the multivariate normal distribution of dimension $k$ with mean
vector $\mu$ and covariance matrix $\Sigma$, and $I\mathcal{G}(\kappa, \lambda)$ an inverse Gamma distribution with shape parameter $\kappa$ and scale
parameter $\lambda$. Provided $X^\top X$ is invertible, this conjugate model leads to closed-form marginal posteriors
\citep{sup:marin:robert:2014}
\begin{align*}
\beta_1, \beta_2 \mid y &\sim \mathcal{T}_2 \left( \frac{n}{n+1}(X^\top X)^{-1}X^\top y, \frac{3+ y^\top(Id - X(X^\top X)^{-1}X^\top)y / 2  }{4+n/2} \frac{n}{n+1} (X^\top X)^{-1}, 8+n \right),\\
\sigma^2 \mid y &\sim I\mathcal{G}\left( 4+\frac{n}{2}, 3+\frac{1}{2}y^\top(Id - X(X^\top X)^{-1}X^\top)y \right),
\end{align*}
where $\mathcal{T}_k\left( \mu, \Sigma, \nu \right)$ is the multivariate Student distribution of dimension $k$, with
location parameter $\mu$, scale matrix $\Sigma$ and degree of freedom $\nu$.

In our simulation experiment, we concentrate on the non zero covariance of the posterior distribution namely $\mbox{Cov}(\beta_1, \beta_2 \mid y)$. A \textit{reference table} of $N=10\,000$ replicates of a $n$-sample with $n=100$ is generated. We
then create $k=60$ summary statistics: the maximum likelihood estimates of $\beta_1$, $\beta_2$, the residual sum of
squares, the empirical covariance 
and correlation between $y$ and $x_1$, covariance and correlation between $y$ and $x_2$, the sample mean, the sample
variance, the sample median, 
and 50 independent noise variables simulated from a uniform distribution $\mathcal{U}_{[0,1]}$. These noise variables were introduced to be in a sparse context.

Similarly to the Normal example of the main text, we assess the performance of our approach using an independent (Monte Carlo)
test dataset of size $N_{\text{pred}}=100$ and compare estimation accuracy with the ABC-RF approach from 
the ones with adjusted ridge regression and neural network ABC methodologies. RF are once again built with $B=500$ trees, $n_{try}=k/3$ and minimum node size equals to $5$ and ABC methods rely on the {\sf R} package \texttt{abc} with a tolerance parameter equals to $0.1$ for ABC methods with adjustment. ABC with neural network adjustment require the specification of the number of layers composing the neural network. We use again 10 layers, the default number of layers in the {\sf R} package \texttt{abc}. For local linear or ridge regression the corrections are univariate.
That is not the case for neural networks which, by construction, perform
multivariate correction.

\begin{figure}
  \centering
	\includegraphics[height=6cm]{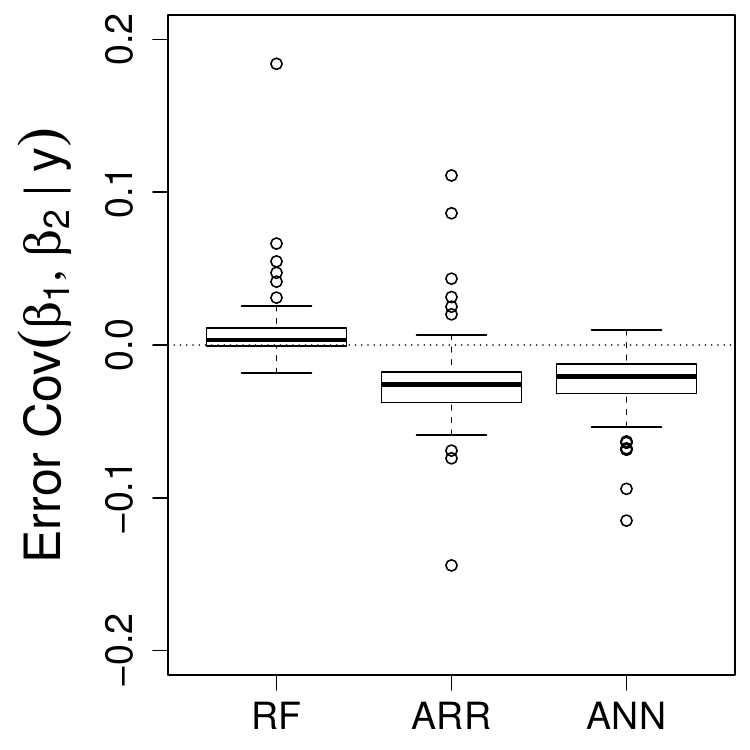}
  \caption{Boxplot comparison of differences between prediction and true values for ${\mbox{Cov}(\beta_1, \beta_2 \mid y)}$ using random forest (RF), adjusted ridge regression (ARR) and adjusted neural network (ANN) ABC methods.}
  \label{fig:QQplot-covariances}
\end{figure}

\begin{figure}[H]
  \centering
        \includegraphics[width=0.75\linewidth]{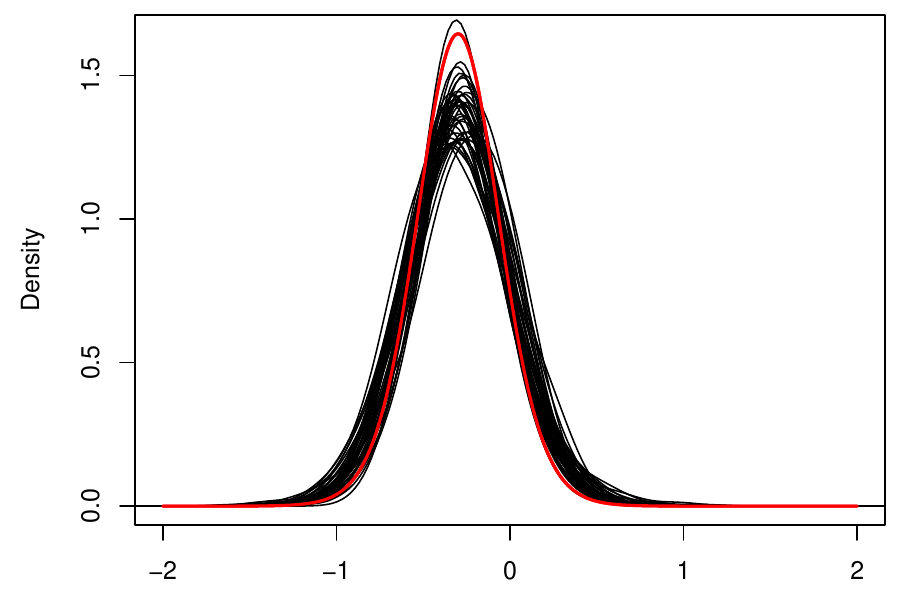}
        \includegraphics[width=0.75\linewidth]{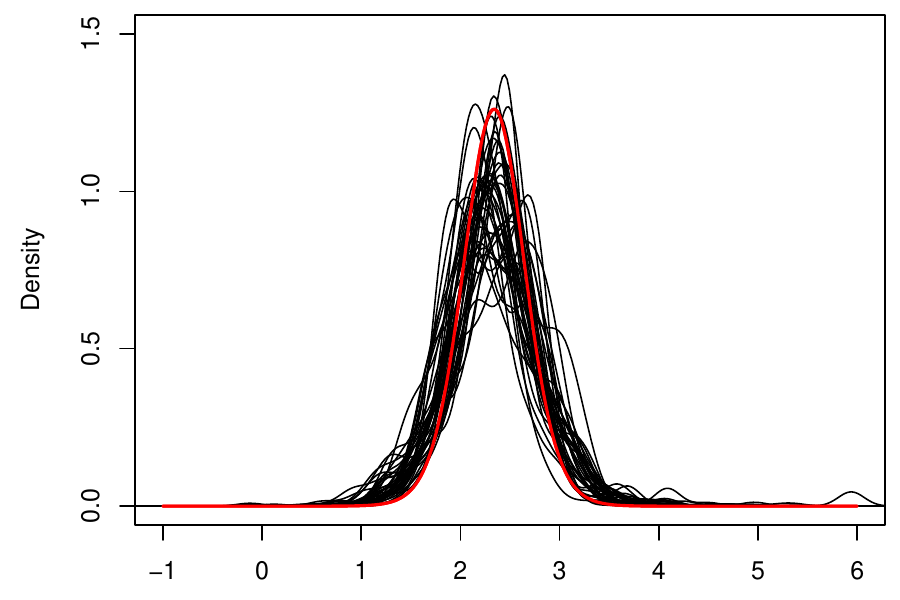}
  \caption{ Comparisons of the true posterior density distribution function of $\theta_1$ in the Normal
model with a sample of $40$ ABC-RF approximations of the posterior density (using RF weights), based on $40$
independent {\textit{reference tables}} and for two different test datasets
(upper and lower panels). True posterior densities are represented by
red lines. }
  \label{fig:true_vs_RFcdfs}
\end{figure}

\begin{figure}[H]
  \centering
  \includegraphics[width=0.5\linewidth]{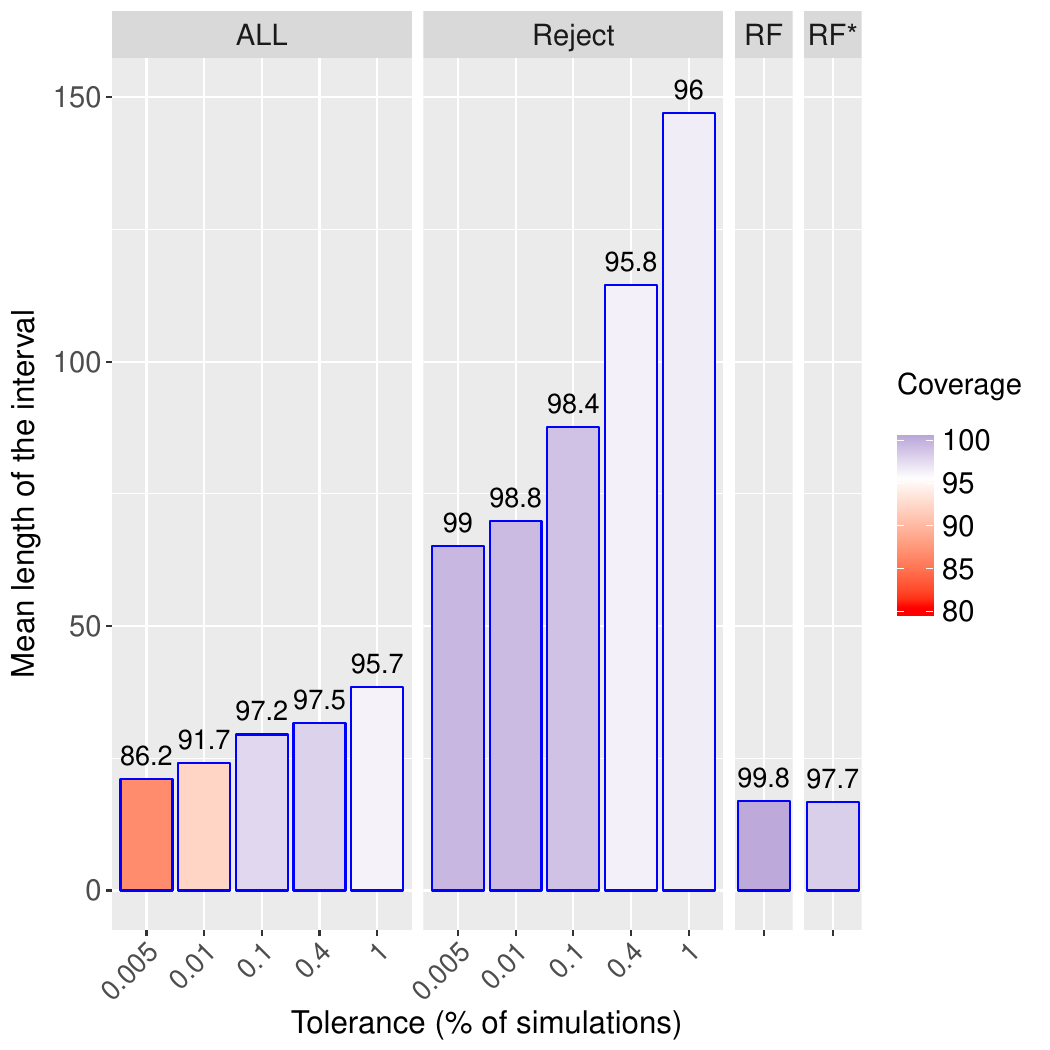}%
  \includegraphics[width=0.5\linewidth]{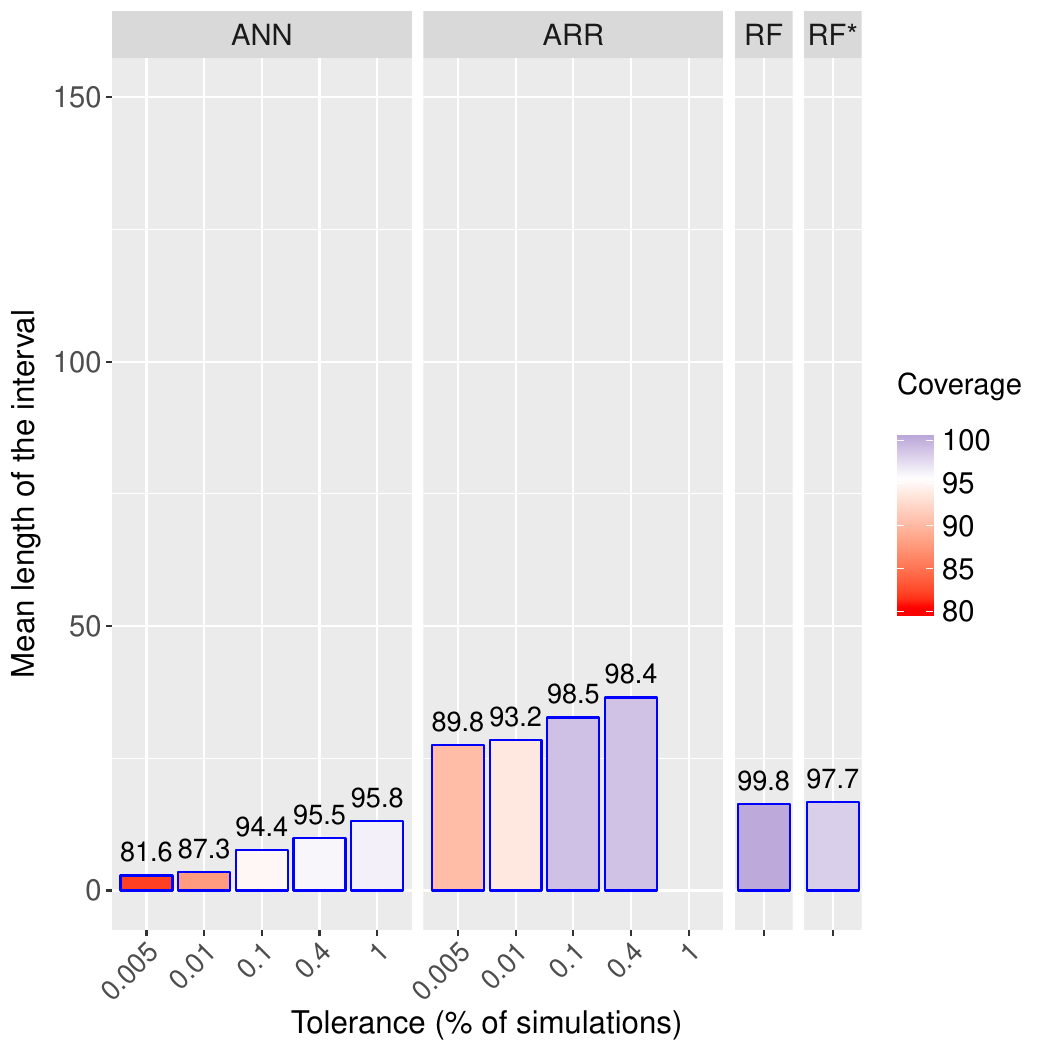}
  \caption{Range and coverage comparison of approximate $95\%$ credible
intervals on the ratio N2/Na of the Human population genetics example,
obtained with ABC-RF (RF) and with earlier ABC methods : rejection (Reject), adjusted local linear (ALL) or ridge regression (ARR) or neural network (ANN) with various tolerance levels for Reject, ALL, ARR and ANN. Coverages values are specified by bar colors
and superimposed values. Heights indicate CI mean lengths. Na is the ancestral African effective population size before the
population size change event and N2 the African effective population size after
the population size change event (going backward in time). RF$^*$ refers to results obtained using ABC-RF when adding 20 additional independent noise variables generated from a uniform $\mathcal{U}_{[0,1]}$ distribution. RF refers to results without noise variables.}
    \label{fig:human:N2Na}
\end{figure}

Covariance estimation is a novel feature in this example, Table~\ref{tab:reg-NMAE} shows that the ABC-RF approach does better in NMAE terms. As exhibited in Figure~\ref{fig:QQplot-covariances}, ABC-RF overestimates covariances when earlier ABC methods underestimate it. Results are quite encouraging even though we believe the method might still be improved.

\begin{table}
\centering
\begin{tabular}{c c c c}
 & RF & ARR & ANN \\
\hline 
$\mbox{Cov}(\beta_1, \beta_2 \mid y)$ & \textbf{0.26} & 0.85 & 0.64 \\
\end{tabular}
\caption{ \label{tab:reg-NMAE} Comparison of normalized mean absolute errors (NMAE) of estimate posterior covariances between $\beta_1$ and $\beta_2$ using random forest (RF), adjusted ridge regression (ARR) and adjusted neural network (ANN) ABC methods. The smallest NMAE value is in bold characters.}
\end{table}
\vspace{2cm}

\clearpage

\section{A basic {\sf R} code to use the \texttt{abcrf} package version 1.7}
\label{sup:sec:codeR}

We provide some basic {\sf R} lines of code to use the {\sf R} package \texttt{abcrf} and conduct RF inference about parameters. There are two possibilities to read simulated data: the user wants to use a \textit{reference table} simulated from the software DIYABC v.2.1.0 \citep{cornuet:etal:2014} recorded within a \texttt{.bin} file associated with its \texttt{.txt} header file, or the simulated \textit{reference table} is only contained within a \texttt{.txt} file. Of course if the model is simple enough, the user can simulate the \textit{reference table} himself using its own simulator program.
In the following, we assume $\theta$ is a vector of \texttt{p} parameters and \texttt{k} summary statistics are considered.
The $\#$ symbol means the text on its right is a comment and ignored by {\sf R}. We here focus on a single parameter of interest labelled ``\texttt{poi}''.

\subsection*{Installing and loading the {\sf R} package \texttt{abcrf}}

\begin{verbatim}
install.packages("abcrf") # To install the abcrf package (version 1.7)
library(abcrf) # To load the package.
\end{verbatim} 

\subsection*{Reading data: option 1 - using a \texttt{.bin} and \texttt{.text} files obtained using DIYABC}

\textcolor{blue}{We assume the \textit{reference table} is recorded within the \texttt{reftable.bin} file and its corresponding header in the \texttt{header.txt} file. The function \texttt{readRefTable} is used to recover the data.}

\begin{verbatim}
data <- readRefTable(filename = "reftable.bin", header = "header.txt")
# data is a list containing the scenarios (or models) indices, the matrix 
# with the parameters, the summary statistics and other informations.

# We are here interested in the simulated data of the scenario 1.
index1 <- data$scenarios == 1 # To store the model 1 indexes.


# We then create a data frame composed of the parameter of interest poi and
# the summary statistics of the scenario 1.
data.poi <- data.frame(poi = data$params[index1, "poi"], 
                       sumsta = data$stats[index1, ])
\end{verbatim}

\subsection*{Reading data: option 2 - using a \texttt{.txt} file}

\textcolor{blue}{We assume that the \textit{reference table} is recorded within \texttt{yourTxtFile.txt} file, composed of a first column corresponding to the scenario indices, \texttt{p} columns of parameters and \texttt{k} columns of summary statistics, the first row is the column labels. The field separator character being a white space.}

\begin{verbatim}
data <- read.table(file = "youTxtFile.txt", header = TRUE, sep = "")
# data is a matrix. The first column is the model indices, the next p are 
# the p parameters, the last k are the summary statistics.

index1 <- data[ , 1] == 1 # To store the model 1 indexes.

# We then create a data frame composed of the parameter of interest poi and
# the summary statistics of model 1. p and k have to be defined.
data.poi <- data.frame(poi = data[index1, "poi"], 
                       sumsta = data[index1, (p+2):(p+k+1)])
\end{verbatim}

\subsection*{Subsetting your dataset}

\textcolor{blue}{If required, subsetting your datasets stored in \texttt{data.poi} can be easily done with the following line.}
\begin{verbatim}
data.poi <- data.poi[1:10000, ]
# If you are interest in the 10000 first datasets.
\end{verbatim}

\subsection*{Training a random forest}

\textcolor{blue}{The random forest of the ABC-RF method is built thanks to the \texttt{regAbcrf} function, its principle arguments being a {\sf R} formula and the corresponding data frame as training dataset. Additional arguments are available, especially the number of trees (\texttt{ntree}, with default values \texttt{ntree} = 500), the minimum node size (\texttt{min.node.size}, with default value \texttt{min.node.size} = 5), and the number of covariates randomly considered at each split (\texttt{mtry}). See the \texttt{regAbcrf} help for further details.}
\begin{verbatim}
model.poi <- regAbcrf(formula = poi~., data = data.poi, ntree = 500,
                      min.node.size = 5, paral = TRUE)
# The used formula means that we are interested in explaining the parameter 
# poi thanks to all the remaining columns of data.poi (i.e. all the 
# summary statistics).
# The paral argument determine if parallel computing will be activated 
# or not.
\end{verbatim}

\subsection*{Graphical representations to access the performance of the method}

\textcolor{blue}{The evolution of the out-of-bag mean squared error depending on the number of tree can be easily represented with the \texttt{err.regAbcrf} function (e.g. Figure 6 of the main text).}

\begin{verbatim}
errorOOB <- err.regAbcrf(object = model.poi, training = data.poi,
                         paral = TRUE)
\end{verbatim}
\textcolor{blue}{The contribution of summary statistics in ABC-RF estimation for the parameter of interest can be retrieved with the \texttt{plot} function applied to an object resulting from \texttt{regAbcrf}.}

\begin{verbatim}
plot(x = model.poi, n.var = 25) 
# The contributions of the 25 most important summary statistics are 
# represented (e.g. Figure S5).
\end{verbatim}

\subsection*{Making predictions}

\textcolor{blue}{Finally, given a data frame \texttt{obs.poi} containing the summary statistics you want the predictions of posterior quantities of interest for a given dataset (usually the observed dataset). When using DIYABC, note that the summary statistics of the observed dataset are recorded in a file name \texttt{statobs.txt}.
The \texttt{predict} method can be used for this purpose. The column names need to be the same than those in the summary statistics of \texttt{data.poi}. }

\begin{verbatim}
# Reading the observed dataset with
obs.poi <- read.table("statobs.txt", skip=2)
# If obs.poi is not a dataframe or the column names do not match, 
# you can use the following lines:
obs.poi <- as.data.frame(obs.poi)
colnames(obs.poi) <- colnames(data.poi[ ,-1])

# Prediction is complete by
pred.obsPoi <- predict(object = model.poi, obs = obs.poi, 
                       training = data.poi, quantiles = c(0.025,0.975),
                       paral = TRUE)
# The 2.5 and 97.5 order quantiles are computed by specifying 
# quantiles = c(0.025,0.975).

# pred.obsPoi is a list containing predictions of interest.

# Posterior mean can be retrieved by
pred.obsPoi$expectation

# Posterior variance by
pred.obsPoi$variance

# Posterior quantiles by
pred.obsPoi$quantiles
\end{verbatim}

\textcolor{blue}{A graphical representation of the approximate posterior density of \texttt{poi} given \texttt{obs.poi}
can be obtained using the \texttt{densityPlot} function.}

\begin{verbatim}
densityPlot(object = model.poi, obs = obs.poi, training = data.poi, paral = TRUE)
\end{verbatim}

\clearpage

\section{Supplementary figures for the Normal toy example}
\label{sup:sec:figureEcdf}

\clearpage

\section{Summary statistics available in the software DIYABC v.2.1.0 for SNP data}
\label{sec:SS}

For single nucleotide polymorphic (SNP) markers, the program DIYABC v.2.1.0 \citep{cornuet:etal:2014} proposes a set of summary statistics among those used by population geneticists. These summary statistics are mean values, variance values and proportion of null values across loci, which allow a rough description of the allelic spectrum. Such summary statistics characterize a single, a pair or a trio of population samples.\\\\
\vspace{0.25cm} \noindent Single population statistics \\
\verb+HP0_i+: proportion of monomorphic loci for population i \\
\verb+HM1_i+: mean gene diversity across polymorphic loci \citep{nei:1987} \\
\verb+HV1_i+: variance of gene diversity across polymorphic loci \\
\verb+HMO_i+: mean gene diversity across all loci \citep{nei:1987} \\\\
\vspace{0.25cm} \noindent Two population statistics \\
\verb+FP0_i&j+: proportion of loci with null FST distance between the two samples for populations i and j \citep{weir:cockerham:1984} \\ 
\verb+FM1_i&j+: mean across loci of non null FST distances  \\
\verb+FV1_i&j+: variance across loci of non null FST distances  \\
\verb+FMO_i&j+: mean across loci of FST distances \citep{weir:cockerham:1984} \\
\verb+NP0_i&j+: proportion of 1 loci with null Nei's distance  \citep{nei:1972}  \\
\verb+NM1_i&j+: mean across loci of non null Nei's distances  \\
\verb+NV1_i&j+: variance across loci of non null Nei's distances  \\
\verb+NMO_i&j+: mean across loci of Nei's distances \citep{nei:1972} \\\\
\vspace{0.25cm} \noindent Three population statistics \\
\verb+AP0_i_j&k+: proportion of loci with null admixture estimate when pop. i comes from an admixture between j and k \\
\verb+AM1_i_j&k+: mean across loci of non null admixture estimate \\
\verb+AV1_i_j&k+: variance across loci of non null admixture estimated \\
\verb+AMO_i_j&k+: mean across all locus admixture estimates \citep{choisy:etal:2004}

\clearpage

\section{Supplementary figures for the Human population genetics example}
\label{sup:sec:figureHuman}

\clearpage

\section{Contribution of summary statistics in ABC-RF estimation of the parameters ra and N2/Na of the Human population genetics example}
\label{sup:importance}

In the same spirit than in \citet{sup:pudlo:etal:2016}, a by-product of our ABC-RF-based approach is to automatically determine the (most) relevant statistics for the estimation of each parameter by computing a criterion of variable importance (here a variable is a summary statistic). For a summary statistic $j$, this measure is equal to the total amount of decrease of the residual sum of squares (RSS) due to splits over this given summary statistic, divided by the number of trees.

Indeed, at a given parent node where $j$ is used, with $n$ datasets, and for a given parameter of interest $\tau$, the decrease of the RSS due to the split event is defined by the formula
\begin{align*}
\sum_{i=1}^{n}(\tau_i-\bar{\tau})^2 - \left( \sum_{i \in \text{ left node}}(\tau_i-\bar{\tau}_L)^2 \quad + \sum_{i \in \text{ right node}}(\tau_i-\bar{\tau}_R)^2 \right),
\end{align*}
where $\bar{\tau}$, $\bar{\tau}_L$ and $\bar{\tau}_R$ are respectively the average of parameter values of the datasets in the parent node, left daughter and right daughter nodes.
Computing this decrease among all nodes where a summary statistics $j$ is used, among all the trees of the forest and dividing it by the number of trees is an importance measurement of the covariate $j$.

\begin{figure}
  \centering
  \includegraphics[width=0.47\linewidth]{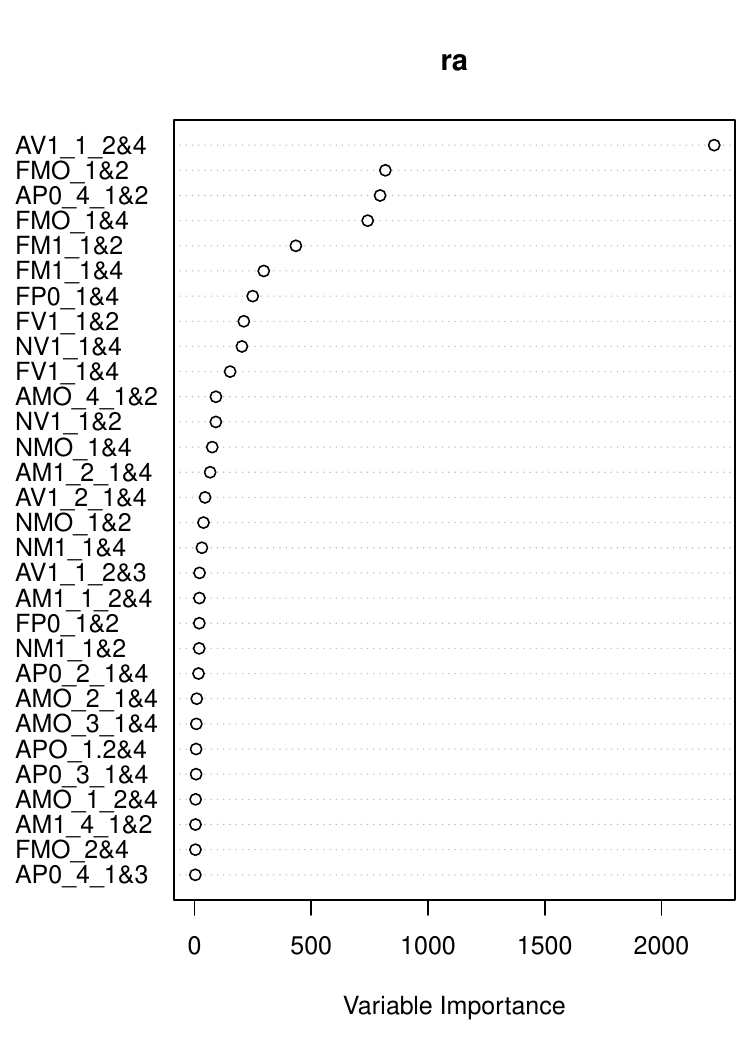} \hspace{0.5cm}
  \includegraphics[width=0.47\linewidth]{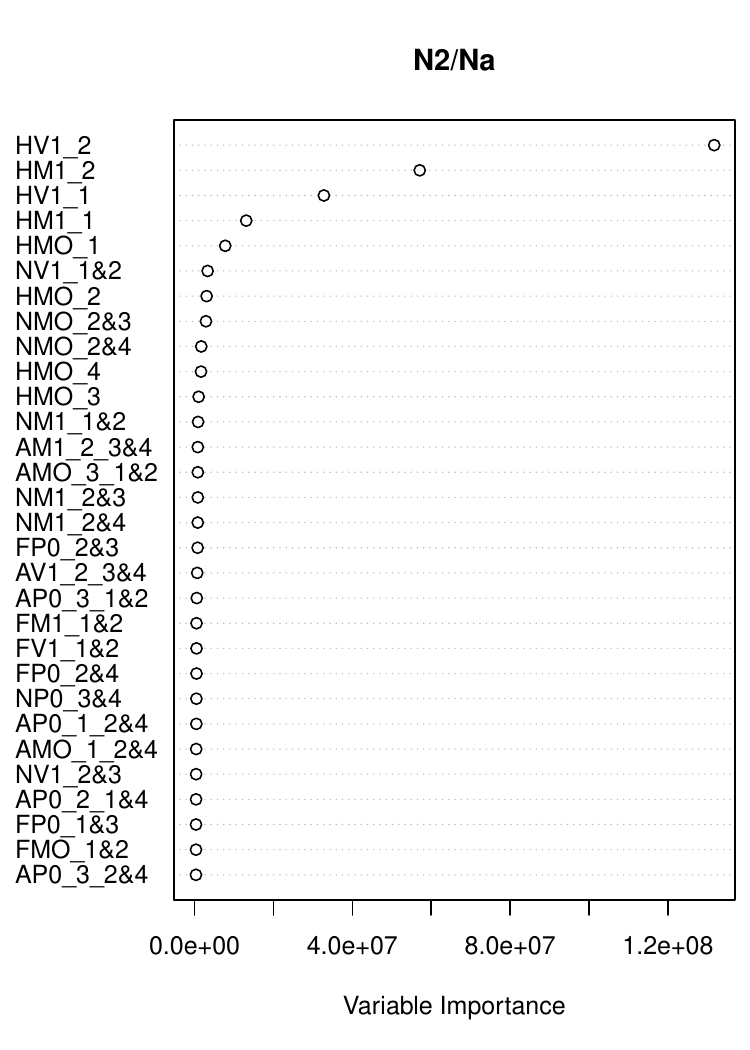}
  \caption{Contributions of the 30 most important summary statistics for the ABC-RF estimation of the parameters ra and N2/Na of the Human population genetics example. The contribution of each statistics is evaluated as the total amount of decrease of the residual sum of squares, divided by the number of trees, for each of the 112 used summary statistics provided for SNP markers by DIYABC. The higher the variable importance the more informative the statistics. The population index(s) is indicated at the end of each statistics. $1=$ pop ASW (Americans of African ancestry), $2=$ pop YRI (Yoruba, Africa), $3=$ pop CHB (Han, Asia) and $4=$ pop GBP (British, Europe). For instance FMO\_1\&4 $=$ mean across loci of Fst distance between the populations 1 and 4. See also Section \ref{sec:SS} of the supplementary materials and the Figure 4 of the main text. Note the difference of scale for the importance criterion for parameters ra and N2/Na. This difference can be explained by the difference of scale in parameter values. Indeed, it directly influences the RSS. The parameter ra being bounded in $[0,1]$ contrary to N2/Na, a higher decrease can be expected for the ratio N2/Na than ra.}
  \label{fig:importance_var}
\end{figure}

Figure \ref{fig:importance_var} shows the contributions of the 30 most important summary statistics (among the 112 statistics proposed by DIYABC) for the ABC-RF estimation of the parameters ra and N2/Na of the Human population genetics example (see Section 3.2 of the main text). The most informative summary statistics are clearly different depending on the parameter of interest. For the admixture rate between two sources populations (ra), all ten most informative statistics correspond to statistics characterizing a pair or a trio of populations (e.g. AV or FMO statistics; see Section \ref{sec:SS} of the supplementary materials). Moreover, all those ``best" statistics include the populations ASW, GBP and YRI which correspond to the target and the two source populations respectively. On the contrary, for the effective population size ratio N2/Na, seven of the ten most most informative statistics correspond to statistics characterizing within population genetic variation (e.g. HV or HMO; see Section \ref{sec:SS} of the supplementary materials). In this case, all those ``best" statistics include the African population, which makes sense since N2 is the effective population size in the studied African population and Na in the population ancestral to all studied populations. It is worth stressing that, although the most informative summary statistics make sense in relation to the studied parameters it was difficult if not impossible to a priori and objectively select those statistics. This is not an issue when using the ABC-RF approach as the method automatically extracts the maximum of information from the entire set of proposed statistics.

\clearpage

\section{Computation times required by the statistical treatments of the studied methods processed following the generation of the reference table}
\label{sup:temps}

We here present a comparison of the computation time requirement for the different methods studied in this paper, when predicting estimations of the admixture rate ra in the human population genetics example. ABC methods with rejection or adjusted with local linear regression provide the best results in terms of CPU time even when the tolerance level is equal to 1. The ABC-RF strategy requires moderately higher computing time. The calculation of the RF weights is the most expensive computation part (i.e. 3/4 of computation time). ABC methods using ridge regression or neural network correction become very time consuming when the tolerance level is high. 
\begin{table}[h]
\centerline{
\begin{tabular}{l c c}
Method & Tol. level & CPU time (in minutes) \\
\hline
RF & NA & 16.64  \\
Reject & 0.005 & 7.54 \\ 
Reject & 0.01 &  7.54 \\
Reject & 0.1 & 7.78 \\
Reject & 0.4 &  7.98 \\
Reject & 1 &  9.14 \\
ALL & 0.005 & 7.66 \\ 
ALL & 0.01 & 7.70 \\
ALL & 0.1 & 8.81 \\
ALL & 0.4 & 9.21 \\
ALL & 1 & 11.32 \\
ARR & 0.005 & 6.71 \\ 
ARR & 0.01 & 6.97 \\
ARR & 0.1 & 40.57 \\
ARR & 0.4 & 560.39 \\
ARR & 1 & $-$  \\
ANN & 0.005 & 22.31 \\
ANN & 0.01 & 33.60 \\
ANN & 0.1 & 216.61 \\
ANN & 0.4 & 1160.67 \\
ANN & 1 &  4028.63 \\
\end{tabular}
}
\caption{\label{tab:human:time} Comparison of the computation time (in minutes) required - after the gereration of the reference table - for the estimation of the parameter of interest ra on a dataset test table, using ABC-RF (RF), ABC with rejection (Reject), adjusted local linear (ALL), ridge regression (ARR) and neural network (ANN), with various tolerance levels for Reject, ALL, ARR and ANN. The test table included 1000 pseudo-observed datasets and the reference table included $199\,000$ simulated datasets summarized with 112 statistics. Results were computed on a cluster with 28 CPU cores of 2.4 GHz. NA stands for not appropriate. }
\end{table}

\end{document}